\documentstyle[12pt,epsf]{article}
\def\hybrid{\topmargin -20pt    \oddsidemargin 0pt
        \headheight 0pt \headsep 0pt
        \textwidth 6.5in        
        \textheight 9in         
        \marginparwidth .875in
        \parskip 5pt plus 1pt   \jot = 1.5ex}

\hybrid
\makeatletter
\@addtoreset{equation}{section}
\makeatother

\newcommand{\cA}{{\cal A}}

\newcommand{\cD}{{\cal D}}

\newcommand{\cL}{{\cal L}}
\newcommand{\cM}{{\cal M}}
\newcommand{\cN}{{\cal N}}

\newcommand{\cR}{{\cal R}}

\newcommand{\hf}{\frac12}
\newcommand{\qt}{\frac14}
\newcommand{\bea}{\begin{eqnarray}}
\newcommand{\eea}{\end{eqnarray}}
\newcommand{\be}{\begin{equation}}
\newcommand{\ee}{\end{equation}}
\newcommand{\bt}{\begin{tabular}}
\newcommand{\et}{\end{tabular}}
\newcommand{\ba}{\begin{array}}
\newcommand{\ea}{\end{array}}

\newcommand{\vev}[1]{\langle #1 \rangle}
\newcommand{\intmod}[1]{\left\lceil{#1}\right\rceil}

\newcommand{\tr}{\mathop{\rm tr}}

\newcommand{\diag}{{\rm diag}}
\newcommand{\R}{{\rm Re}}
\newcommand{\I}{{\rm Im}}

\def \RR{\hbox{\rm I}\!\hbox{\rm R}}

\def \mnorm#1{{\textstyle{#1}}}

\def \jb{{\bar\jmath}}

\def \half{{\textstyle\hf}}

\def \quarter{{\textstyle\qt}}
\def \soll={\stackrel{!}{=}}
\def \rf=#1{\stackrel{(\ref{#1})}{=}}
\def \bfm#1{\mbox{\boldmath$#1$}}

\def \ith{\int d^2\theta\;}
\def \ithb{\int d^2\bar\theta\;}
\def \ithth{\int d^2\theta d^2\bar\theta\;}
\def \Z#1{$\bfm{Z}_{#1}$}
\def \bZ{{\mathbf Z}}

\def \w{{\rm with\ }}

\def \N#1{$\cN=#1$}
\def \D#1{$D=#1$}
\def \mn{_{\mu\nu}}
\def \MN{^{\mu\nu}}
\def \dm{\partial_\mu}
\def \dn{\partial_\nu}
\def \dM{\partial^\mu}
\def \dN{\partial^\nu}
\def \eps{\epsilon_{\mu\nu\rho\sigma}}
\def \Eps{\epsilon^{\mu\nu\rho\sigma}}
 
\def \barfill{\leaders\hrule height 0.1 true pt\hfill}
\def \overbar#1{\vbox{\ialign{##\crcr\barfill\crcr\noalign{\kern 1pt
                                      \nointerlineskip}$\hfil{#1}\hfil$\crcr}}}
\def \scriptbar#1{{\vbox{\ialign{##\crcr\thinspace\barfill\thinspace\crcr
    \noalign{\kern 0.8pt\nointerlineskip}$\hfil{\scriptstyle #1}\hfil$\crcr}}}}

\def \myvrule#1{\rule[-2.5ex]{0.1mm}{5.5ex}{}\hskip0.5mm
                 \raisebox{-1.5ex}{${}_{#1}$}}
\def \myvruletwo#1#2{\rule[-3.5ex]{0.1mm}{7ex}{}\hskip0.5mm
            \raisebox{-1.5ex}{${}_{{\scriptstyle#1}\atop{\scriptstyle#2}}$}}

\newlength{\oldindent}
\newlength{\quadlength} 
\newlength{\abstand} \newlength{\breite}

\def \Nucl#1{{\em Nucl.~Phys.}\ {\bf B#1}}
\def \NuclProc#1{{\em Nucl.~Phys.~Proc.~Suppl.}\ {\bf #1}}
\def \PhysR#1{{\em Phys.~Rev.}\ {\bf D#1}}

\def \PhysL#1{{\em Phys.~Lett.}\ {\bf #1B}}
\def \CoMaPhy#1{{\em Comm.~Math.~Phys.}\ {\bf #1}}

\def \JGeom#1{{\em J.~Geom.~Phys.}\ {\bf #1}}
\def \Ann#1{{\em Ann.~Phys.}\ {\bf #1}} 
\def \CQG#1{{\em Class.~Quant.~Grav.}\ {\bf #1}}
\def \jhep#1{{\em JHEP}\ {\bf #1}}
\def \hep#1{{\tt hep-th/#1}}
\def \hepp#1{{\tt hep-ph/#1}}

\def \nonpert{non\discretionary{-}{}{-}per\-tur\-ba\-tive\ }

\def \nonab{non\discretionary{-}{}{-}Abel\-ian\ }
\def \nonvan{non\discretionary{-}{}{-}van\-ish\-ing\ }

\def \bia{b^{\prime i}_a}
\def \aik{\alpha^i_k}
\def \aikt{\tilde\alpha^i_k}

\renewcommand{\thefootnote}{\fnsymbol{footnote}}

\begin{document}

\begin{titlepage}
\begin{center}

\rightline{FTUAM-99/34}
\rightline{IFT-UAM/CSIC-99-39}
\rightline{hep-th/9910143}

\vskip .6in
{\LARGE \bf Anomaly cancellation in\\[1ex]
\D4, \N1 orientifolds\\[1ex]
and linear/chiral multiplet duality}
\vskip .8in

{\bf Matthias Klein\footnote{E-mail: Matthias.Klein@uam.es}\\
\vskip 0.8cm
{\em Departamento de F\'\i sica Te\'orica C-XI and 
     Instituto de F\'\i sica Te\'orica C-XVI \\
     Universidad Aut\'onoma de Madrid, Cantoblanco,
     28049 Madrid, Spain}
}

\end{center}

\vskip 1.5cm

\begin{center} {\bf ABSTRACT } \end{center}
It has been proposed that gauge and K\"ahler anomalies in four-dimensional
type IIB orientifolds are cancelled by a generalized Green-Schwarz mechanism 
involving exchange of twisted RR-fields. We explain how this can be understood
using the well-known duality between linear and chiral multiplets. We find 
that all the twisted fields associated to the \N1 sectors and some of the 
fields associated to the \N2 sectors reside in linear multiplets. But there 
are no linear multiplets associated to order-two twists. Only the linear
multiplets contribute to anomaly cancellation. This suffices to cancel
all $U(1)$ anomalies. In the case of K\"ahler symmetries the complete
$SL(2,\RR)$ can be restored at the quantum level for all planes that are
not fixed by an order-two twist.

\vfill
October 1999
\end{titlepage}

\renewcommand{\thefootnote}{\arabic{footnote}}
\setcounter{footnote}{0}

\section{Introduction}
Anomaly freedom is one of the basic requirements for gauge theories.
At first sight, a gauge theory can only be consistent if the charged
matter content is precisely such that all anomalous one-loop diagrams
vanish. However, in low-energy effective theories of string theory
there is another possibility. The anomalous one-loop diagram can be
cancelled by an additional interaction which in the effective field
theory appears as the tree-level exchange of a field coupling to the
gauge fields in exactly the right manner. In the seminal work of Green
and Schwarz \cite{GS} it is shown how the hexagon anomalies in \D{10}
are cancelled by the exchange of the NSNS 2-form $B$. In \D4 $B$ is 
dual to a scalar $\phi$ which has axionic couplings. The special
interactions between $\phi$ and the gauge fields indeed cancel the
triangle anomalies \cite{DSW}. Only $U(1)$ anomalies that are universal
with respect to the different gauge group factors $G_a$ (i.e.\ the
coefficient of the $G_a^2U(1)$ triangle diagram is independent of $a$)
can be cancelled by this mechanism. Incidentally, all gauge anomalies
in four-dimensional heterotic vacua are of this form.

The situation is different in type I vacua where non-universal $U(1)$
gauge anomalies appear. As string theory is believed to be consistent,
there must be additional interactions that cancel these anomalies. In
\cite{Sagn92,BLPSSW} a generalized version of the Green-Schwarz mechanism 
was proposed which cancels the $U(1)$ anomalies in \D6 via exchange 
of twisted RR fields (explicit calculations for many models are
performed in \cite{SS99}). The authors of \cite{IRU98} applied this
idea to type IIB orientifold vacua in \D4 and showed that all anomalies
can be cancelled this way. They also pointed out that anomalies of
the invariance of the effective theory under K\"ahler transformations
can be cancelled by the same mechanism \cite{IRU99}.

In the present article we explain how the anomalous transformation of
the twisted RR axions, which are responsible for anomaly cancellation, 
can be understood by considering the Chern-Simons couplings of the RR 
2-forms. In the next section we show that the RR 2-forms which appear in the 
twisted spectrum of type IIB orientifolds belong to \N1 linear multiplets%
\footnote{Some useful facts about linear multiplets are assembled in
appendix A; see also \cite{GGRS,deren94,grimm98}.} $L^{(k)}$, where $k$ labels 
the twisted sectors. We find that all sectors except the $k=N/2$ sector (for
even $N$) contain linear multiplets. The two crucial 
ingredients for this generalized Green-Schwarz mechanism to work are
a Chern-Simons modification of the field strength associated to the
RR 2-form and an additional term $\ithth \sum_k L^{(k)} \tr(\gamma_k V)$
in the \D4 effective Lagrangian ($V$ is the vector multiplet
corresponding to the anomalous $U(1)$ and $\gamma_k$ is the matrix
representing the twist on the gauge indices). Both terms can be derived
from string theory by considering either ``branes within branes'' \cite{BinB}
or by using the inflow mechanism \cite{GHM} (for \D6 the relevant couplings
are calculated in \cite{MSS98}).

A linear multiplet $L^{(k)}$ is dual to a chiral multiplet $M^{(k)}$ in
the sense that there exists an equivalent description of the theory
in terms of $M^{(k)}$ \cite{lin_chir,grimm98}. Translating the 
Chern-Simons modification of the RR field strength and the $L^{(k)}V$
term to the chiral basis, one finds that the $M^{(k)}$ couple to the
gauge fields just in the right way to cancel the anomalies. This is
discussed in section 3.

In section 4 we apply this method to K\"ahler anomalies and argue that
a coupling of the form $\ithth \sum_{i,k}\aik L^{(k)}\ln(T_i+\bar T_i)$
must be present, where $T_i$ is the chiral superfield which parameterizes 
the K\"ahler class of the $i$-th complex compact plane. This leads
to a non-trivial transformation of the dual chiral multiplets of twisted
states $M^{(k)}$ and cancels the K\"ahler anomalies. More precisely,
only the K\"ahler symmetries corresponding to complex planes that are not
fixed under an order-two twist can be restored at the quantum level by
this mechanism. However, an order-two twist implies the existence of
D5-branes. The coupling of the $T_3$ (we chose the third complex plane
to be fixed under the order-two twist) to the D5-brane gauge fields 
explicitly breaks the K\"ahler symmetry for the third plane. Therefore 
one cannot expect an anomaly cancellation for this plane.
In contrast to the situation in heterotic string vacua, the full $SL(2,\RR)$
symmetry seems to be preserved for fixed planes which are not fixed under
an order-two twist.

In section 5 we give the low-energy effective Lagrangian describing the
interactions of the moduli (dilaton, twisted and untwisted moduli) and
the gauge fields. We determine the values of the Fayet-Iliopoulos terms and
the masses of the (pseudo-)anomalous gauge bosons.

An appendix summarizes useful results of \N1 supersymmetry, concerning linear 
multiplets and D-terms of general chiral Lagrangians.

\section{Massless twisted spectrum of type IIB orientifolds}
In this article we consider compact \Z{N} orientifolds (see \cite{GJ,AFIV} 
and references therein). They are obtained from
the ten-dimensional type IIB theory by compactifying on a six-dimensional torus
and projecting onto \Z{N}- and $\Omega^\prime$-invariant states. Here 
$\Omega^\prime=\Omega J$, with $\Omega$ the world-sheet parity and $J$ an
operator acting on the twisted sectors \cite{Pol96} as explained below.
For consistency one has to add twisted closed strings and open strings. 
The compact dimensions form a six-torus $T^6=\RR^6/\Lambda$, where $\Lambda$
is a six-dimensional lattice chosen such that the generator $\theta$ of \Z{N}
acts on it as an automorphism. Acting on $\Lambda$, $\theta$ is a $(6\times6)$
integer matrix. Its action on the coordinates of $T^6$ can be diagonalized
over the complex numbers: 
$\theta=\diag(e^{2\pi iv_1},e^{2\pi iv_2},e^{2\pi iv_3})$, where we grouped
the coordinates of $T^6$ into three complex planes. The twist vector
$v=(v_1,v_2,v_3)$ is chosen such that $0<|v_i|<1$ and $\sum_i v_i=0$.
In general there are different lattices $\Lambda$, on which a twist 
$\theta(v)$ acts as an automorphism. The orbifold is defined to be the space
$T^6/\bZ_N$, whereas the orientifold includes an additional projection 
of the strings moving on the orbifold onto $\Omega^\prime$-invariant states.
All possible \Z{N}-twists $\theta$ and lattices $\Lambda$ leading to 
\D4, \N1 orbifolds have been classified in \cite{EK}. In \cite{AFIV}
it was found that of the possible \Z{N} models only the ones shown in
table \ref{tzn} lead to consistent type IIB orientifolds 
(i.e.\ the tadpoles can be cancelled).
\begin{table}[htb]
\begin{center}
$$\begin{array}{|c|c||c|c|}
\hline
\bZ_3  &\frac13(1,1,-2)    &\bZ_6           &\frac16(1,1,-2) \\ \hline
\bZ_7  &\frac17(1,2,-3)    &\bZ_6^{\prime}  &\frac16(1,-3,2) \\ \hline
       &                   &\bZ_{12}        &\frac1{12}(1,-5,4) \\
\hline
\end{array}$$
\end{center}
\caption{\label{tzn}\Z{N} actions in \D4.}
\end{table}

The massless spectrum of these models has been determined in \cite{AFIV}.
However in the twisted sector only the number of states is given. In addition,
for each twist $\theta(v)$ only one possible lattice (implicitly assumed as
factorizable) is considered.
As we are interested in the precise \N1 multiplets containing the twisted
states, we rederive the spectrum by considering the cohomology of the 
compact orbifold space \cite{EK,Zas,VW}. This method was used in 
\cite{DP96,Zwa} to obtain the massless spectrum of 
twisted states of orientifold models in \D6 and \D4. 
The orbifold cohomology is encoded in the Hodge numbers
$h^{p,q}=\dim H^{p,q}$, where $H^{p,q}$ is the space of harmonic 
$(p,q)$-forms. For six-dimensional compact orbifolds (i.e.\ \D4 non-compact
dimensions) one has
\be \label{hnum}
h^{0,0}=h^{3,0}=h^{0,3}=h^{3,3}=1, 
\qquad h^{1,1}=h^{2,2}, \qquad h^{2,1}=h^{1,2}.
\ee
Here $h^{1,1}$ and $h^{2,1}$ are the numbers of \Z{N}-invariant $(1,1)$-forms
and $(2,1)$-forms of the six-torus. There are additional contributions 
$h^{1,1}_{\rm tw}$ and $h^{2,1}_{\rm tw}$ from the twisted sector.

The bosonic fields of type IIB theory in \D{10} are the dilaton $\varphi$, the
metric $g$ and a 2-form $B$ from the NSNS sector and a scalar $C_{(0)}$,
a 2-form $C_{(2)}$ and a 4-form $C_{(4)}$ from the RR sector. Under the
world-sheet parity $\varphi$, $g$ and $C_{(2)}$ are even, the other fields
being odd. In the untwisted sector $\Omega^\prime=\Omega$ and therefore
the bosonic fields in \D4 are obtained by contracting the Lorentz indices
of $\varphi$, $g$ and $C_{(2)}$ with the harmonic forms of the orbifold.
From $h^{0,0},h^{3,0},h^{0,3},h^{3,3}$ one gets the \D4 graviton, dilaton
and antisymmetric tensor. The latter two fields (together with the 
corresponding fermion) form an \N1 linear multiplet. From the remaining
harmonic forms one finds $h^{1,1}$ chiral multiplets $T_i$ corresponding
to deformations of the K\"ahler class and $h^{2,1}$ chiral multiplets $U_i$
corresponding to deformations of the complex structure.

Let us now consider the twisted sectors. For the $k$-th twisted sector 
there is a singular subspace $\cM_k$ which is fixed under the action of
$\theta^k$. The cohomology $H^{p,q}_{k\hbox{-}\rm twisted}$ of this space 
contributes to $H^{p+n_k,q+n_k}$ of the orbifold, where $n_k=\sum_i {}^kv_i$
and ${}^kv_i=k\,v_i$ mod $\bfm Z$, such that $0\leq{}^kv_i<1$. The spaces
$\cM_1$ and $\cM_{N-1}$ consist of the set of fixed points of the orbifold.
More generally, for twists $\theta^k$, such that all ${}^kv_i\neq0$, the
spaces $\cM_k$ and $\cM_{N-k}$ consist of the set of fixed points under
the action of $\theta^k$. For such $k<N/2$ one finds $n_k=1$, $n_{N-k}=2$
(with two exceptions, explained below, where $n_k=2$, $n_{N-k}=1$).
If ${}^kv_i=0$ for some $i$, then the $i$-th complex plane is fixed
under $\theta^k$. In this case the spaces $\cM_k$ and $\cM_{N-k}$ consist 
of the set of fixed planes. For such $k$ one finds $n_k=n_{N-k}=1$.
Denote the number of $\theta^k$ fixed points by $f_k$ (if ${}^kv_i\neq0$). 
If ${}^kv_i=0$ 
for some $i$ let $f_k^\prime$ be the number of $\theta^k$ fixed planes.
On the other hand we define $f^{(i)}_k$ to be the number of $\theta^k$ 
fixed points of the four-dimensional space consisting only of the two 
rotated planes (i.e.\ the compact space without the $i$-th plane).
If the lattice $\Lambda$ splits into a direct sum (over the integers)
of sublattices, $\Lambda=I\oplus J$, such that I is fixed under $\theta^k$
and this block structure is preserved under $\theta^k$, then $f_k^\prime$ 
is just given by $f^{(i)}_k$.
As noted by the authors of \cite{EK}, this condition is not always satisfied, 
leading (in some cases) to a smaller value of $f_k^\prime$.
However, we will restrict ourselves to lattices satisfying this condition.
These are the ones that were discussed in \cite{AFIV,IRU98,IRU99}.

Write the contribution of the twisted sectors to the cohomology as 
$h^{p,q}_{\rm tw}=\sum_{k=1}^{N-1}h^{p,q}_k$. If the greatest common divisor
of $k$ and $N$, $\gcd(k,N)$, is a prime number and for $k<N/2$ such that all
${}^kv_i\neq0$, we find
\be \label{hodge_fpt}
h^{1,1}_k=h^{2,2}_{N-k}=f_1+{f_k-f_1\over\gcd(k,N)},
\qquad h^{p,q}_k=0 {\rm \ for\ all\ other\ } (p,q).
\ee
If $\gcd(k,N)=pq$, with $p$, $q$ prime numbers (which is only possible
for $N=12$), this is modified to
\be
h^{1,1}_k=h^{2,2}_{N-k}=f_1+{f_p-f_1\over p}+{f_q-f_1\over q}
                        +{f_k-f_1-(f_p-f_1)-(f_q-f_1)\over pq}.
\ee
Here we used the fact that on a point one can only define a $(0,0)$-form.
If $\gcd(k,N)\neq1$, then only $f_1$ of the $f_k$ $\theta^k$ fixed points 
are invariant under \Z{N}. The remaining $(f_k-f_1)$ fixed points transform
under a \Z{\gcd(k,N)} (resp.\ \Z{p} or \Z{q} if $\gcd(k,N)=pq$) subgroup 
of \Z{N}. One can however form linear combinations of $\gcd(k,N)$ fixed 
points that are invariant under the whole \Z{N}. Note that $h^{1,1}_k=0$
for $k>N/2$ and $h^{2,2}_k=0$ for $k<N/2$, with two exceptions. This means
that for $k<N/2$ the $h^{2,2}_{N-k}$ forms furnish the antiparticles
corresponding to the particles from the $h^{1,1}_k$ forms. The two exceptions
are the $k=3$ sector of \Z7 and the $k=5$ sector of \Z{12}, where the 
particles come from the $(N-k)$-th sector (i.e.\ $h^{1,1}_{N-k}\neq0$)
and the antiparticles from $h^{2,2}_k$. 

In the case of fixed planes one has to consider the torus cohomology which
is $h^{0,0}=h^{1,0}=h^{0,1}=h^{1,1}=1$. The $(0,0)$-form and the $(1,1)$-form 
are \Z{N}-invariant but the $(1,0)$-form and the $(0,1)$-form generically
transform under a \Z{k} subgroup of \Z{N}. If ${}^kv_i=0$, then $f^{(i)}_1$
is the number of tori that are fixed under the whole \Z{N}. Again one can 
form linear combinations of the the other $\theta^k$ fixed tori and of 
the forms defined on them that are invariant under the whole \Z{N}. 
In total, one finds for $\gcd(k,N)=pq$ (if $\gcd(k,N)=$prime, set $q=1$)
\bea  \label{hodge_fpl}
&&h^{1,1}_k=h^{2,2}_k=f_1^{(i)}+{f_p^{(i)}-f_1^{(i)}\over p}
                        +{f_q^{(i)}-f_1^{(i)}\over q}
                        +{f_k^{(i)}-f_1^{(i)}-(f_p^{(i)}-f_1^{(i)})
                                   -(f_q^{(i)}-f_1^{(i)})\over pq},
   \nonumber\\
&&h^{2,1}_k=h^{1,2}_k=h^{1,1}_k-f_1^{(i)},\\
&&h^{p,q}_k=0 {\rm \ for\ all\ other\ } (p,q). \nonumber
\eea

For the orbifolds of table \ref{tzn} we find the twisted Hodge numbers 
listed in table \ref{hzn}. 
(For completeness we added the untwisted Hodge numbers in the last column.)
\begin{table}[htb]
$$\begin{array}{|l|l|l|l|}
\hline
\bZ_3  &h^{1,1}_1=27,\ h^{1,1}_2=0
       &h^{1,1}_{\rm tw}=27  &h^{1,1}=9 \\ \hline
\bZ_6  &h^{1,1}_1=3,\ h^{1,1}_2=3+12,\ h^{1,1}_3=1+5,\ h^{1,1}_{4/5}=0
       &h^{1,1}_{\rm tw}=24  &h^{1,1}=5 \\
       &h^{2,1}_3=5,\ h^{2,1}_{1/2/4/5}=0
       &h^{2,1}_{\rm tw}=5 & \\ \hline
\bZ_6^{\prime}  &h^{1,1}_1=12,\ h^{1,1}_2=h^{1,1}_4=3+3,\ h^{1,1}_3=4+4,\ 
                 h^{1,1}_5=0
       &h^{1,1}_{\rm tw}=32   &h^{1,1}=3 \\
       &h^{2,1}_2=h^{2,1}_4=3,\ h^{2,1}_3=4,\ h^{2,1}_{1/5}=0
       &h^{2,1}_{\rm tw}=10  &h^{2,1}=1 \\ \hline
\bZ_7  &h^{1,1}_1=7,\ h^{1,1}_2=7,\ h^{1,1}_4=7,\ h^{1,1}_{3/5/6}=0   
       &h^{1,1}_{\rm tw}=21  &h^{1,1}=3 \\ \hline    
\bZ_{12}  &h^{1,1}_1=3,\ h^{1,1}_2=3,\ h^{1,1}_3=h^{1,1}_9=1+1,\ 
           h^{1,1}_4=3+6 & & \\
       &h^{1,1}_6=1+1+2,\ h^{1,1}_7=3,\  h^{1,1}_{5/8/10/11}=0
       &h^{1,1}_{\rm tw}=26  &h^{1,1}=3 \\
       &h^{2,1}_3=h^{2,1}_9=1,\ h^{2,1}_6=1+2,\ h^{2,1}_{1/2/4/5/7/8/10/11}=0
       &h^{2,1}_{\rm tw}=5 & \\ \hline
\end{array}$$
\caption{\label{hzn}Twisted (and untwisted) Hodge numbers of \Z{N} orbifolds.}
\end{table}

The twisted bosonic fields of the \D4 theory are obtained by contracting
the bosonic fields in \D{10} with the additional harmonic forms from the
twisted sectors. Now the operator $J$ in $\Omega^\prime=\Omega J$ is important
because it exchanges the sector twisted by $\theta^k$ with the one twisted
by $\theta^{N-k}$ \cite{Pol96}. To get an $\Omega^\prime$-invariant result,
one has to contract the $\Omega$-even fields $g$ and $C_{(2)}$ with the
$J$-even linear combinations of harmonic forms from the $k$-th and $(N-k)$-th
twisted sector and the $\Omega$-odd fields $B$ and $C_{(4)}$ with $J$-odd
linear combinations of harmonic forms.

From the twisted sectors with no fixed planes one finds $h^{1,1}_k$ scalars
in \D4 from the $J$-even sector and the same number of antisymmetric tensors 
from the $J$-odd sector. Together with their
fermionic partners they form $h^{1,1}_k$ \N1 linear multiplets. If there 
are fixed planes, one has $2(h^{1,1}_k+h^{2,1}_k)$ scalars from the
$J$-even sector and $h^{1,1}_k$ scalars, $h^{1,1}_k$ antisymmetric tensors
and $h^{2,1}_k$ vectors from the $J$-odd sector. 
Together with the corresponding fermions these fields form 
$(h^{1,1}_k+h^{2,1}_k)$ chiral multiplets, $h^{1,1}_k$ linear
multiplets and $h^{2,1}_k$ vector multiplets if $k\neq N/2$. These fit into
$h^{2,1}_k$ \N2 hyper multiplets (consisting of two \N1 chiral multiplets), 
$h^{2,1}_k$ \N2 vector-tensor multiplets \cite{WKLL} (consisting of an 
\N1 vector and an \N1 linear multiplet) and $(h^{1,1}_k-h^{2,1}_k)$ \N2 
linear hyper multiplets (consisting of an \N1 chiral and an \N1 linear 
multiplet).\footnote{\label{Ntwo}The spectrum of the $k$-th twisted 
sector with ${}^kv_i=0$ can also be understood by compactifying on a 
four-dimensional \Z{m} orientifold \cite{GJ,DP96}, with $m=N/\gcd(k,N)$, 
and then dimensionally reducing on a two-torus. For the $k^\prime$-th 
twisted sector of the \Z{m} orientifold the authors find 
$f_k^{(i)}+(f_{kk^\prime}^{(i)}-f_k^{(i)})/\gcd(k^\prime,m)$
\D6, \N1 hypers and, if $k^\prime\neq m/2$, the same number of 
additional tensors (a $k^\prime\neq m/2$ sector exists if $m\neq2$,
i.e. $k\neq N/2$). 
From these, $f_1^{(i)}=(h^{1,1}_k-h^{2,1}_k)$ correspond to \Z{N} fixed
points in the four-dimensional compact space. The remaining multiplets
can be grouped into $h^{2,1}_k$ different \Z{N}-invariant linear combinations.
From each of the $f_1^{(i)}$ \D6, \N1 hypers and tensors only half the states 
are \Z{N}-invariant (in terms of \D4, \N1: a linear and a chiral multiplet). 
The $h^{2,1}_k$ linear combinations of twisted states
reduce to \N2 hypers and vector-tensors in \D4. (The latter are equivalent
to \N2 vectors because an antisymmetric tensor is dual to a scalar in \D4.)}
For $k=N/2$ there are no $J$-odd linear combination of harmonic forms and 
therefore in this case only the $(h^{1,1}_{N/2}+h^{2,1}_{N/2})$ 
chiral multiplets appear in the \D4 twisted spectrum. 

Adding the contributions from the twisted sectors 
$k=1,\ldots,\intmod{N/2}$ (because of the orientifold 
projection the sectors $k>N/2$ give no independent degrees of freedom)
we find the following twisted fields, table 3.
\begin{table}[htb]
$$\begin{array}{|l|c|c|}
\hline
&\mbox{sectors without} &\mbox{sectors with}\\[-1ex]
&\mbox{fixed planes}    &\mbox{fixed planes} \\ \hline\hline
\bZ_3  &27{\rm\ \ lin.\ mult.} &- \\ \hline
\bZ_6  &18{\rm\ \ lin.\ mult.}
       &11{\rm\ \ chir.\ mult.}\\ \hline
\bZ_6^{\prime}  &12{\rm\ \ lin.\ mult.}
                &21{\rm\ \ chir.\ mult.},\ 6{\rm\ \  lin.\ mult.},\ 
                 3{\rm\ \  vector\ mult.}\\ \hline
\bZ_7  &21{\rm\ \  lin.\ mult.} &- \\ \hline
\bZ_{12}  &18{\rm\ \ lin.\ mult.}
          &10{\rm\ \ chir.\ mult.},\ 2{\rm\ \ lin.\ mult.},\ 
           1{\rm\ \ vector\ mult.}\\ \hline
\end{array}$$
\caption{\label{twzn}Twisted fields (\N1 multiplets) of \Z{N} orientifolds.}
\end{table}

To understand the role of the twisted fields in the anomaly cancellation
mechanism, it is important to note two facts: First, all twisted fields
associated to twists that rotate all planes appear in \N1 linear multiplets.
Second, in the twisted sectors with fixed planes only some of the fields are
in \N1 linear multiplets and all of them appear in the $k\neq N/2$ sectors.
We will see that only \N1 linear multiplets can contribute to anomaly
cancellation.

\section{Gauge anomaly cancellation}
It is well known that the \D4 version of the Green-Schwarz anomaly 
cancellation mechanism \cite{DSW} is closely related to the equivalence
between the linear and the chiral \N1 supermultiplets. More precisely, 
in the heterotic string the transformation of the chiral dilaton
superfield that cancels the anomaly can be understood by starting 
with the linear dilaton multiplet that appears in the string spectrum
and then translating the Lagrangian carefully to the description in terms 
of a chiral field (see e.g.\ \cite{LO,DFKZ,AGNT}). In the present article we 
apply this method to the anomaly cancellation mechanism in \N1, \D4 type 
IIB orientifolds \cite{IRU98,IRU99}. The new features are that several
linear multiplets are involved and that their vacuum expectation value
is not related to the string loop expansion but rather to the blowing-up
of the orbifold singularities.

\subsection{Purely bosonic case}
We start by reviewing the \D4 Green-Schwarz mechanism as it can be derived
from the equivalence between an antisymmetric tensor $B\mn$ and a scalar
$\phi$ without making any use of supersymmetry.\footnote{In this and in the
next subsection we follow the line of reasoning of \cite{grimm98}.}
Start from an auxiliary Lagrangian describing the interactions of a 
non-dynamical field $Y^\mu$ with the antisymmetric tensor and several 
gauge fields:
\bea \label{Lns_aux}
\cL^\prime &= &-{2\over b_1}Y_\mu Y^\mu + Y_\mu\Eps\dn B_{\rho\sigma}
        +2b_2^{(a)}(Y_\mu+b_3A_\mu^{(0)})\tilde Q^{(a)\mu} \nonumber\\
        &&+b_3\tilde F\mn^{(0)}B\MN 
          -{1\over4g_{(a)}^2}\tr(F\mn^{(a)}F^{(a)\mu\nu}).
\eea
Here $b_1$, $b_2^{(a)}$, $b_3$ are dimensionful coupling constants. The 
gauge fields $A_\mu^{(a)}$ correspond to different gauge group factors $G_a$
(which can be Abelian or non-Abelian); a sum over $a$ is understood. 
We choose $G_0=U(1)_X$ to be the (pseudo-)anomalous 
symmetry. $Q^{(a)}$ is the the Chern-Simons 3-form associated to $G_a$:
\be \label{defQ}
Q^{(a)}_{\mu\nu\rho}=\tr\left(A^{(a)}_{[\mu}\dn A^{(a)}_{\rho]}
                  -{2i\over3}A^{(a)}_{[\mu}A^{(a)}_{\nu}A^{(a)}_{\rho]}\right).
\ee
A tilde denotes the Poincar\'e dual
\be \label{Poindual}
\tilde F^{(a)\mu\nu}=\hf\Eps F^{(a)}_{\rho\sigma},\qquad
\tilde Q^{(a)\mu}={1\over3!}\Eps Q^{(a)}_{\nu\rho\sigma},\qquad
\tilde H^\mu={1\over3!}\Eps H_{\nu\rho\sigma},
\ee
where $H$ is the modified antisymmetric tensor field strength
\be \label{Hdef}
H_{\mu\nu\rho}=\partial_{[\mu}B_{\nu\rho]}+b_2^{(a)}Q^{(a)}_{\mu\nu\rho}.
\ee
Useful identities are
\bea \label{dQ_FF}
  \dm\tilde Q^{(a)\mu} &= &\half\tr(F^{(a)}\mn\tilde F^{(a)\mu\nu}),\\
\label{HH}
  \tilde H^\mu\tilde H_\mu &= &-{1\over3!}H^{\mu\nu\rho}H_{\mu\nu\rho}.
\eea

The Lagrangian (\ref{Lns_aux}) contains no kinetic terms for $Y$ and $B$.
Therefore these fields can be integrated out using the equations of motion.
By varying with respect to $Y$ or $B$ two equivalent (``dual'') descriptions 
are obtained from $\cL^\prime$.

(1) From $\delta_Y\cL^\prime=0$ one finds
\be \label{Y_B}
Y^\mu={b_1\over4}(\Eps\dn B_{\rho\sigma}+2b_2^{(a)}\tilde Q^{(a)\mu})
     =\hf b_1\tilde H^\mu.
\ee
Inserting this into (\ref{Lns_aux}) gives
\be \label{L_B}
\cL^{B}=\half b_1\tilde H_\mu\tilde H^\mu+b_3\tilde F^{(0)}\mn B\MN
        +2b_3b_2^{(a)}\tilde Q^{(a)}_\mu A^{(0)\mu}
        -\mnorm{1\over4g_{(a)}^2}\tr(F^{(a)}\mn F^{(a)\mu\nu}),
\ee
with $H=dB+b_2^{(a)}Q^{(a)}$. This is the usual Lagrangian describing
an antisymmetric tensor coupled to gauge fields via Chern-Simons terms
plus two additional terms proportional to $b_3$. They are crucial to
cancel the $U(1)_X$ anomalies.

(2) If instead one requires $\delta_B\cL^\prime=0$, one finds
$\Eps(-\dn Y_\mu+b_3\dm A_\nu^{(0)})=0$, which is solved by
\be \label{Y_phi}
Y_\mu=\hf\dm\phi-b_3A_\mu^{(0)},
\ee
where $\phi$ is an arbitrary function.
Inserting this into (\ref{Lns_aux}) gives after some simple algebra
\be \label{L_phi}
\cL^{\phi}=-\mnorm{1\over2b_1}(\dm\phi-2b_3A_\mu^{(0)})(\dM\phi-2b_3A^{(0)\mu})
           -\mnorm{1\over4g_{(a)}^2}\tr(F^{(a)}\mn F^{(a)\mu\nu})
           -\half b_2^{(a)}\phi\tr(F^{(a)}\mn\tilde F^{(a)\mu\nu}).
\ee
The theory described by this Lagrangian is equivalent to the one described
by (\ref{L_B}), and in this sense the scalar $\phi$ is dual to the
antisymmetric tensor $B\mn$. The normalization for $\phi$ in (\ref{Y_phi})
is chosen such that a canonically normalized kinetic term for $B\mn$ in
(\ref{L_B}), i.e.\ $b_1=1$, implies a canonically normalized kinetic term
for $\phi$ in (\ref{L_phi}). In our conventions $B\mn$ is dimensionless,
$A_\mu$ and $H_{\mu\nu\rho}$ have dimension of mass, $F\mn$ and $\phi$ 
have dimension of mass squared, $Q_{\mu\nu\rho}$ and $Y_\mu$ have 
dimension of mass cubed. This implies that the coefficients $b_1$ and 
$b_3$ have dimension of mass squared and $b_2$ has dimension of inverse 
mass squared.

The first term in (\ref{L_phi}) tells us that $\phi$ transforms 
non-trivially under $U(1)_X$ gauge transformations:
\be \label{phi_transf}
\delta_\epsilon A_\mu^{(0)}=\dm\epsilon \qquad\Rightarrow\qquad
\delta_\epsilon\phi=2b_3\epsilon.
\ee
One observes that the presence of the Chern-Simons term in the antisymmetric
tensor field strength $H=dB+b_2^{(a)}Q^{(a)}$ in (\ref{L_B}) leads to the
$\phi\tr(F\tilde F)$ term in (\ref{L_phi}), and the $b_3\tilde F^{(0)}B$
term in (\ref{L_B}) leads to the transformation law (\ref{phi_transf}).

The variation of $\cL^\phi$ under a $U(1)_X$ transformation is given by
\be \label{del_L}
\delta_\epsilon\cL^\phi=\left({\cA^{(a)}\over32\pi^2}\epsilon
 -\hf b_2^{(a)}\delta_\epsilon\phi\right)\tr(F^{(a)}\mn\tilde F^{(a)\mu\nu}).
\ee
Here $\cA^{(a)}$ is the coefficient of the $(G_a)^2U(1)_X$ triangle diagram,
$\cA^{(a)}=\sum_{\cR}q_{\cR}T(\cR)$, where the sum goes over all fields
transforming in representations $\cR$ under $G^{(a)}$ and carrying $U(1)_X$
charge $q_{\cR}$. The generators $\lambda^i_{(a)}$ of $G_a$ are normalized to
$\tr(\lambda^i_{(a)}\lambda^j_{(a)})=\delta^{ij}$.

Using (\ref{phi_transf}) one sees that all $U(1)_X$ anomalies are cancelled
if
\be \label{cancel_cond}
{\cA^{(a)}\over32\pi^2}=b_2^{(a)}b_3.
\ee

In the heterotic string in \D{10} the Chern-Simons modification of the
antisymmetric tensor field strength is fixed by the supergravity algebra
\cite{BRWN} and leads to an almost universal coefficient 
$b_2^{(a)}=-k^{(a)}\alpha^\prime/4$ in \D4, depending only on the Kac-Moody 
level $k^{(a)}$. The explicit string calculation \cite{FIcalc} shows that 
at one loop the terms proportional to $b_3$ in (\ref{L_B}) are indeed 
generated and that the coefficient is 
$b_3=-\cA^{(a)}/(k^{(a)}8\pi^2\alpha^\prime)$, 
independently of $a$, which is just the right value to cancel the
anomalies.

For the case of type IIB orientifolds we need to generalize the above
formulae to include several antisymmetric tensors $B\mn^{(k,f)}$, where
$k$ labels the different twisted sectors $k=1,\ldots,\intmod{N-1\over2}$
and $f=1,\ldots,f_k$ counts the fixed points\footnote{Here we restrict 
ourselves to the the antisymmetric tensors from the twisted sectors with 
no fixed planes. We will see that only in one case, namely the second twisted
sector of $\bZ_6^\prime$, fields associated to fixed planes contribute to 
anomaly cancellation.} of each twisted sector (in general not all of these 
fields are independent). The $k=N/2$ sector (for even $N$) is excluded,
because we saw in the previous section that it contains no antisymmetric
tensors. Only certain linear combinations of the antisymmetric tensors
are involved in the anomaly cancellation mechanism. We define
\be \label{def_Bk}
B\mn^{(k)}={1\over\sqrt{f_k}}\sum_f B\mn^{(k,f)}.
\ee

From the study of RR forms and their coupling to D-branes it has been
found that in the presence of gauge fields on the D-branes the RR field
strengths are modified by Chern-Simons terms (see e.g.\ \cite{BinB,MSS98}).
In our case of twisted RR 2-forms we have
\be \label{twHdef}
H^{(k,f)}_{\mu\nu\rho}=\partial_{[\mu}B_{\nu\rho]}^{(k,f)}
        +\epsilon_kc_2\sum_a Q^{(k,a)}_{\mu\nu\rho}\qquad\Rightarrow\qquad
H^{(k)}_{\mu\nu\rho}=\partial_{[\mu}B_{\nu\rho]}^{(k)}
        +\epsilon_kc_2\sum_a \sqrt{f_k}Q^{(k,a)}_{\mu\nu\rho},
\ee
where $c_2$ is a normalization factor of mass dimension $-2$ and $\epsilon_k$
is a sign, that will be determined below.
The Chern-Simons form now depends on the twist matrix $\gamma_k$ that 
represents the action of $\theta^k$ on the gauge indices,
\be
Q^{(k,a)}_{\mu\nu\rho}=\tr\left(\gamma_k\left(A^{(a)}_{[\mu}\dn A^{(a)}_{\rho]}
       -{2i\over3}A^{(a)}_{[\mu}A^{(a)}_{\nu}A^{(a)}_{\rho]}\right)\right).
\ee

The study of RR forms also shows \cite{BinB,MSS98} that terms of the form
$\tr(\gamma_k\exp(iF))\wedge C$ appear in the effective action, 
where $F$ is the gauge field strength 2-form, $C$ is the sum over RR 
forms of different degrees and it is understood that after expansion 
of the exponential only the terms with the correct total form degree 
are kept. 
For us of interest is the term coupling the RR 2-form to the 
(pseudo-)anomalous $U(1)_X$\,\footnote{For simplicity we assume that only
one $U(1)$ factor is anomalous. The generalization to several anomalous
$U(1)$'s is straightforward.}:
\be \label{BF_term}
c_3\tr(\gamma_ki\tilde F^{(0)}\mn)B^{(k)\mu\nu},
\ee
where $c_3$ is a normalization factor of mass dimension 2.

In the conventions of \cite{AFIV} the $\gamma_k$'s are parametrized by
a 16-dimensional shift vector $V$ and a 16-dimensional vector $H$ containing
the Cartan generators of $SO(32)$: $\gamma_k=\exp(-2\pi ikH\cdot V)$.
The trace over the product of $\gamma_k$ with the gauge group generators
$\lambda_{(a)}$ is easily calculated to give \cite{IRU98}:
$\tr(\gamma_k\lambda_{(a)}^2)=\cos(2\pi kV_a)$ and
$\tr(\gamma_ki\lambda_{(0)})=2n_X\sin(2\pi kV_X)$. Here $V_a$ is the component
of the shift vector $V$ corresponding to the gauge group $G_a$ and $n_X$ is
the rank of the non-Abelian group in which $U(1)_X$ is embedded. Comparison
with (\ref{L_B}) where we normalized the traces to one leads to the
identification
\be \label{bcoeff}
    b_2^{(k,a)}=\epsilon_kc_2\sqrt{f_k}\cos(2\pi kV_a), \qquad 
    b_3^{(k)}=2n_Xc_3\sin(2\pi kV_X).
\ee

In direct generalization of the equivalence between the Lagrangians
(\ref{L_B}) and (\ref{L_phi}), we find that the effective theory
describing the RR 2-forms,
\bea \label{L_Bk}
\cL^{B} &= &\sum_k\left(\half b_1\tilde H^{(k)}_\mu\tilde H^{(k)\mu}
        +b_3^{(k)}\tilde F^{(0)}\mn B^{(k)\mu\nu}
        +2b_3^{(k)}b_2^{(k,a)}\tilde Q^{(a)\mu}A^{(0)}_{\mu}\right) \nonumber\\
  &&    -\mnorm{1\over4g_{(a)}^2}\tr(F^{(a)}\mn F^{(a)\mu\nu}),
\eea
where $Q^{(a)}$ (in contrast to $Q^{(k,a)}$ in (\ref{twHdef})) is defined 
and normalized as in (\ref{defQ}), can equivalently be described in terms 
of scalars $\phi^{(k)}$,
\bea \label{L_phik}
\cL^{\phi} &= &-\mnorm{1\over2b_1}\sum_k(\dm\phi^{(k)}-2b_3^{(k)}A_\mu^{(0)})
                     (\dM\phi^{(k)}-2b_3^{(k)}A^{(0)\mu})  \nonumber \\
 &&  -\mnorm{1\over4g_{(a)}^2}\tr(F^{(a)}\mn F^{(a)\mu\nu})
     -\sum_k\half b_2^{(k,a)}\phi^{(k)}
     \tr(F^{(a)}\mn\tilde F^{(a)\mu\nu}).
\eea
Here and in the following the sum over the twisted sectors runs over
$k=1,\ldots,\intmod{N-1\over2}$.

It is easy to see that all $U(1)_X$ anomalies are cancelled if
\be \label{cancel}
{\cA^{(a)}\over32\pi^2}=\sum_kb_2^{(k,a)}b_3^{(k)}.
\ee
For \D4, \N1 type IIB orientifolds the anomaly coefficient $\cA^{(a)}$ 
introduced above can be calculated as a function of the shift vector $V$ 
and is given by 
\cite{IRU98}\footnote{A factor 2 compared to the result of \cite{IRU98} is
due to our choice of normalization of gauge group generators, 
$\tr(\lambda^i\lambda^j)=\delta^{ij}$.}
\be \label{anom_coeff}
\cA^{(a)}={-4\over N}\sum_kC_kn_X\sin(2\pi kV_X)\cos(2\pi kV_a),
\ee
where $C_k=\prod_{i=1}^32\sin(\pi kv_i)$ if the gauge groups $G_a$ and 
$U(1)_X$ live either both on the D9-branes or both on the D5-branes 
\cite{IRU98}. In this case one has $(C_k)^2=f_k$ if the $k$-th sector 
has no fixed planes and $C_k=0$ else. An explicit calculation of the various
$C_k$'s for the considered models shows that $C_k=-\sqrt{f_k}$ for the sectors
$k<N/2$ of all models, except the $k=3$ sector of \Z7 and the $k=5$ sector
of \Z{12}. This is related to the fact, that these two sectors furnish
antiparticles, whereas in all other cases the antiparticles come from
$k>N/2$, as we saw in the previous section. Let us define $\epsilon_k=1$
for the third sector of \Z7 and the fifth sector of \Z{12} and
$\epsilon_k=-1$ for all other sectors. Then we have
\be \label{Ck_fk} 
    C_k=\epsilon_k\sqrt{f_k}. 
\ee
If $G_a$ and $U(1)_X$ live on different types of branes (this is called
the 95-sector), then $C_k=2\sin(\pi kv_3)$, where it was assumed that 
the D5-branes are extended in the four-dimensional space-time and in the 
third complex plane and are at the origin in the first and second complex 
planes. It turns out that the only contribution to $\cA^{(a)}$ from 
twisted sectors with fixed planes is from the 95-sector and is only 
in the $k=2$ sector of $\bZ_6^\prime$ non-vanishing. In the previous 
section we saw that there are six antisymmetric tensors from this sector 
associated to the second plane. But only three of them can couple to the 
D5-branes because they are associated to fixed planes located at the origin 
in the first complex direction. In analogy to (\ref{def_Bk}) we therefore 
define $B\mn^{(2)}=(1/\sqrt3)\sum_{f=1}^3 B\mn^{(2,f)}$ (and $\epsilon_2=1$).
In the $k=3$ sector of \Z{12} there are two more antisymmetric tensors,
associated to the third complex plane. They do not seem to couple to the
D5-branes. From (\ref{bcoeff}) one therefore finds that all gauge anomalies 
are cancelled if
\be \label{ccond}
c_2c_3={-1\over16\pi^2N}.
\ee
It seems difficult to obtain the separate values of the coefficients $c_2$
and $c_3$ because, under a field redefinition $B\mn^{(k)}\to\alpha B^{(k)}\mn$,
these coefficients scale as $c_2\to\alpha c_2$ and $c_3\to c_3/\alpha$.
If one chooses a normalization such that the term in front of
$\quarter\tr(F\tilde F)$ in (\ref{L_phik}) is 
\be 
{\alpha^\prime\over N}\sum_{k,f}\phi^{(k,f)}
         \epsilon_k\tr(\gamma_k\lambda^2_{(a)})
={\alpha^\prime\over N}\sum_k\phi^{(k)}\epsilon_k\sqrt{f_k}\cos(2\pi kV_a),
\ee
then $c_2=\alpha^\prime/2N$. The string tension $\alpha^\prime$ is inferred
from dimensional analysis (recall that $\phi$ has dimension of mass squared). 
A careful determination of the coefficient
appearing in (\ref{BF_term}) should then yield 
$c_3=-1/(8\pi^2\alpha^\prime)$ to satisfy the condition (\ref{ccond}).

\subsection{Supersymmetric case}
The supersymmetric generalization of the auxiliary Lagrangian (\ref{Lns_aux})
is
\be \label{L_aux}
\cL^\prime=\ithth\left({1\over4b_1}Y^2+YL+4b_3V^{(0)}L\right),
\ee
where $Y$ is a real but otherwise unconstrained superfield, $V^{(0)}$
is the vector multiplet corresponding to the $U(1)_X$ gauge symmetry
and $L$ is a (modified) linear multiplet containing the antisymmetric
tensor field strength. In components one has (see eqs.\ 
(\ref{Lcomp}), (\ref{defOmega}) of appendix A)
\bea \label{comp_L}
L &= &l+\theta\chi+\bar\theta\bar\chi+\theta\sigma^\mu\bar\theta\tilde H_\mu
   +\ldots, \\
&&\w \tilde H_\mu=\half\eps\dN B^{\rho\sigma}+b_2^{(a)}\tilde Q^{(a)}_\mu.
  \nonumber
\eea
It satisfies the constraints
\be \label{constr_L}
\cD^2L=2b_2^{(a)}\tr(\bar W^{(a)}\bar W^{(a)}),\qquad
\bar\cD^2L=2b_2^{(a)}\tr(W^{(a)}W^{(a)}),
\ee
where $\cD^2=\cD^\alpha\cD_\alpha$, 
$\bar\cD^2=\bar\cD_{\dot\alpha}\bar\cD^{\dot\alpha}$ are the
supercovariant derivatives and $W^{(a)}_\alpha$ is the chiral field
strength multiplet associated to the gauge symmetry $G_a$.
These equations are the supersymmetric generalization of
the modified Bianchi identity 
$\dM\tilde H_\mu=\half b_2^{(a)}\tr(F^{(a)}\tilde F^{(a)})$,
which follows from (\ref{dQ_FF}).

Again two equivalent descriptions can be obtained from (\ref{L_aux})
by varying with respect to $Y$ or $L$.

(1) From $\delta_Y\cL^\prime=0$ one finds $Y=-2b_1L$ and, inserting
this into (\ref{L_aux}),
\be \label{L_L}
\cL^L=\ithth\left(-b_1L^2+4b_3V^{(0)}L\right),
\ee
with $\cD^2L=2b_2^{(a)}\tr(\bar W^{(a)}\bar W^{(a)})$,
$\bar\cD^2L=2b_2^{(a)}\tr(W^{(a)}W^{(a)})$. Expanding this Lagrangian
in component fields (see eqs.\ (\ref{K_of_Lmod}), (\ref{VL_comp})), one 
finds that the bosonic part coincides with the Lagrangian (\ref{L_B}) 
if we set the additional boson $l$ to a constant value $\vev l$ and 
identify the gauge coupling constants as $g_{(a)}^{-2}=2b_1b_2^{(a)}\vev l$.

(2) The variation with respect to $L$ is more subtle because it is
a constrained superfield, satisfying (\ref{constr_L}). A modified
linear multiplet can always be written in terms of unconstrained
superfield spinors $\xi^\alpha$, $\bar\xi_{\dot\alpha}$ and a
3-form multiplet $\Omega$ containing the Chern-Simons 3-form
(see e.g.\ \cite{grimm98,deren94}), such that $L-\Omega$ contains 
no couplings to gauge fields and satisfies the usual linear constraint 
$\cD^2(L-\Omega)=\bar\cD^2(L-\Omega)=0$:
\be \label{def_xi}
L=\cD^\alpha\bar\cD^2\xi_\alpha+\bar\cD_{\dot\alpha}\cD^2\bar\xi^{\dot\alpha}
  +\Omega,
\ee
with $\cD^2\Omega=2b_2^{(a)}\tr(\bar W^{(a)}\bar W^{(a)})$,
$\bar\cD^2\Omega=2b_2^{(a)}\tr(W^{(a)}W^{(a)})$.
An integration by parts and variation with respect to $\xi$, $\bar\xi$
leads to
\be
\bar\cD^2\cD_\alpha(Y+4b_3V^{(0)})=0=\cD^2\bar\cD_{\dot\alpha}(Y+4b_3V^{(0)}),
\ee
which is solved by
\be \label{Y_S}
Y=S+\bar S-4b_3V^{(0)},
\ee
where $S$ is an arbitrary chiral superfield, with components
$S=s+\sqrt2\theta\psi+\theta\theta F+i\theta\sigma^\mu\bar\theta\dm s+
{i\over2}\sqrt2\theta\theta\bar\theta\bar\sigma^\mu\dm\psi+\qt
\theta\theta\bar\theta\bar\theta\Box s$.
If we identify $\I(s)=\phi$, then (\ref{Y_S}) reads 
$Y=\ldots-4\theta\sigma^\mu\bar\theta(\half\dm\phi-b_3A^{(0)}_\mu)+\ldots$, 
which is the generalization of (\ref{Y_phi}). The normalization is chosen 
such that the Lagrangian shown below has canonical kinetic terms if $b_1=1$.
The sign of $S+\bar S$ in (\ref{Y_S}) cannot be fixed; our choice is analogy
to the low-energy effective action of the heterotic string (see below).
Inserting (\ref{Y_S}) into (\ref{L_aux}) and using the identity 
$\ithb SL=-\quarter S\bar\cD^2L=-\half b_2^{(a)}S\tr(W^{(a)}W^{(a)})$, 
one finds
\bea \label{L_S}
\cL^S &= &\ithth{1\over4b_1}\left(S+\bar S-4b_3V^{(0)}\right)^2 \nonumber\\
   &&   -\hf\ith b_2^{(a)}S\tr(W^{(a)}W^{(a)})
        -\hf\ithb b_2^{(a)}\bar S\tr(\bar W^{(a)}\bar W^{(a)}).
\eea
The bosonic part of this Lagrangian coincides with (\ref{L_phi}) if we set
the additional boson $\R(s)$ to a constant value and identify 
$g_{(a)}^{-2}=-2b_2^{(a)}\vev{\R(s)}$ (for consistency one must have
$b_2^{(a)}\vev{\R(s)}<0$).

In the low-energy effective theory of string theory one has a logarithmic
K\"ahler potential for the dilaton superfield. We therefore introduce
the auxiliary Lagrangian
\be \label{Lx}
\cL^{\prime\prime}=\ithth\left(-b_1\ln(Y/b_1)+YL_{\rm dil}
                              +4b_3V^{(0)}L_{\rm dil}\right).
\ee
In exactly the same manner as above one derives the two equivalent 
descriptions.

(1) From $\delta_Y\cL^{\prime\prime}=0$ one finds $Y=b_1/L_{\rm dil}$, 
which (using $\ithth b_1=0$) leads to
\be \label{L_lnL}
\cL^{L_{\rm dil}}=\ithth\left(b_1\ln(L_{\rm dil})
                              +4b_3V^{(0)}L_{\rm dil}\right),
\ee
with $\cD^2L_{\rm dil}=2b_2^{(a)}\tr(\bar W^{(a)}\bar W^{(a)})$,
$\bar\cD^2L_{\rm dil}=2b_2^{(a)}\tr(W^{(a)}W^{(a)})$. An expansion in
components (see eqs.\ (\ref{K_of_Lmod}), (\ref{VL_comp}))
shows that (in the limit where gravity decouples) this is
part of the effective Lagrangian for heterotic string vacua in \D4 with
gauge group $G=\prod_a G_a$. It describes the NSNS 2-form $B\mn$, the
dilaton $\varphi$ in the lowest component of $L_{\rm dil}$,
$l_{\rm dil}=e^{2\varphi}$, and the gauge fields.

(2) Writing $L_{\rm dil}$ again as in (\ref{def_xi}) and varying with respect 
to $\xi$, $\bar\xi$, one finds $Y=S_{\rm dil}+\bar S_{\rm dil}-4b_3V^{(0)}$,
as in eq.\ (\ref{Y_S}). Note that the sign of $S_{\rm dil}+\bar S_{\rm dil}$
in $Y$ is fixed because we want to identify $S_{\rm dil}$ 
with the chiral dilaton multiplet, which implies $\R{\vev s}>0$ and
$b_2^{(a)}<0$. Inserting $Y$ in (\ref{Lx}) yields
\bea \label{L_lnS}
\cL^{S_{\rm dil}} &= &-\ithth b_1\ln\left(S_{\rm dil}+\bar S_{\rm dil}
                                          -4b_3V^{(0)}\right) \nonumber\\
   &&   -\hf\ith b_2^{(a)}S_{\rm dil}\tr(W^{(a)}W^{(a)})
        -\hf\ithb b_2^{(a)}\bar S_{\rm dil}\tr(\bar W^{(a)}\bar W^{(a)}).
\eea
This dual description of heterotic vacua is more familiar because it is
easier to deal with a chiral superfield than with a linear multiplet.
As mentioned above (in the paragraph below (\ref{cancel_cond})), in the 
heterotic string the coefficient $b_2^{(a)}$ is universal up to the 
Kac-Moody level $k^{(a)}$: $b_2^{(a)}=-k^{(a)}\alpha^\prime/4$. We identify 
${\alpha^\prime\over2}s=e^{-2\varphi}+ia$, where $\varphi$
is the dilaton and $a$ the scalar field dual to the antisymmetric tensor.
This leads to gauge couplings $g_{(a)}=e^{\vev\varphi}/\sqrt{k^{(a)}}$.

From the kinetic terms for $S_{\rm dil}$ one sees that, under a $U(1)_X$ gauge
transformation $V^{(0)}\ \to\ V^{(0)}+\hf(\Lambda+\bar\Lambda)$,
$S_{\rm dil}$ transforms as $S_{\rm dil}\ \to \ S_{\rm dil}+2b_3\Lambda$. 
The variation of the Lagrangian is thus given by
\be
\delta_\Lambda\cL^{S_{\rm dil}}=\ith\Lambda\left({\cA^{(a)}\over32\pi^2}
                       -b_2^{(a)}b_3\right) \tr(W^{(a)}W^{(a)})\ +\ {\rm h.c.}
\ee
This vanishes if the condition (\ref{cancel_cond}) is satisfied, which,
as we saw above, does indeed happen for heterotic string vacua in \D4.

In type IIB orientifolds one has several linear multiplets $L^{(k,f)}$
from the twisted sectors, where $k=1,\ldots,\intmod{N-1\over2}$, 
$f=1,\ldots,f_k$ (not all of them are independent). In generalization
of (\ref{def_Bk}) we define the linear combinations
\be L^{(k)}={1\over\sqrt{f_k}}\sum_f L^{(k,f)}.
\ee

The supersymmetric generalization of the Chern-Simons coupling
(\ref{twHdef}) reads
\be \label{twLconstr}
\cD^2L^{(k)}=2c_2C_k\tr(\gamma_k\bar W^{(a)}\bar W^{(a)}),\qquad
\bar\cD^2L^{(k)}=2c_2C_k\tr(\gamma_kW^{(a)}W^{(a)}).
\ee
Here $C_k=\prod_{i=1}^32\sin(\pi kv_i)$ for all sectors except the 
second sector of $\bZ_6^\prime$, where $C_k=\sqrt3$.
The coupling (\ref{BF_term}) of the linear multiplet to the anomalous
$U(1)_X$ takes the form
\be \label{VL_term}
4c_3\tr(i\gamma_kV^{(0)})L^{(k)}.
\ee
Again the traces over the product of $\gamma_k$ with the gauge group generators
can be calculated and lead to the identification
$b_2^{(k,a)}=c_2C_k\cos(2\pi kV_a)$,
$b_3^{(k)}=2n_Xc_3\sin(2\pi kV_X)$, exactly as in the purely
bosonic case, eq.\ (\ref{bcoeff}).

Assuming a quadratic K\"ahler potential for the twisted fields, we find,
in direct generalization of (\ref{L_L}) and (\ref{L_S}), that the effective 
theory describing the linear multiplets
\be \label{L_Lk}
\cL^L=\ithth\sum_k\left(-b_1(L^{(k)})^2+4b_3^{(k)}V^{(0)}L^{(k)}\right),
\ee
with $\cD^2L^{(k)}=2b_2^{(k,a)}\tr(\bar W^{(a)}\bar W^{(a)})$,
$\bar\cD^2L^{(k)}=2b_2^{(k,a)}\tr(W^{(a)}W^{(a)})$, can equivalently
be described in terms of chiral superfields $M^{(k)}$:
\bea \label{L_Mk}
\cL^M &= &\ithth{1\over4b_1}\sum_k\left(M^{(k)}+\bar M^{(k)}
                               -4b_3^{(k)}V^{(0)}\right)^2 \\
   &&   -\hf\ith \sum_kb_2^{(a,k)}M^{(k)}\tr(W^{(a)}W^{(a)})
        -\hf\ithb \sum_kb_2^{(a,k)}\bar M^{(k)}\tr(\bar W^{(a)}\bar W^{(a)}).
   \nonumber
\eea

From the transformation law under $U(1)_X$ gauge transformations,
\be \label{M_transf} 
V^{(0)}\ \to\ V^{(0)}+\hf(\Lambda+\bar\Lambda)\qquad\Rightarrow\qquad
M^{(k)}\ \to\ M^{(k)}\ +2b_3^{(k)}\Lambda,
\ee
one sees that the variation of the Lagrangian is
\be
\delta_\Lambda\cL^{M}=\ith\Lambda\left({\cA^{(a)}\over32\pi^2}
         -\sum_kb_2^{(k,a)}b_3^{(k)}\right) \tr(W^{(a)}W^{(a)})\ +\ {\rm h.c.}
\ee
The condition for gauge anomaly cancellation is therefore the same that was
found in the purely bosonic case (\ref{cancel}).

Let us now analyze the question which twisted fields can contribute to 
anomaly cancellation. The twisted sectors with no fixed planes are commonly
denoted as \N1 sectors and the ones with fixed planes as \N2 sectors. 
As we saw in section 2, all the fields from the \N1 sectors are in \N1 
linear multiplets. The mechanism described above works well for
the fields from the \N1 sectors. 
The twisted fields of the \N2 sectors fit into \N1 chiral and \N2 hyper, 
linear hyper and vector-tensor multiplets. We can think of 
the \N2 sectors as follows. If the $k$-th twisted sector has a fixed plane,
then $\gcd(k,N)\neq1$. Consider the \Z{m} orientifold, with 
$m=N/\gcd(k,N)$, generated by $\theta^k$. This leads to \N2 supersymmetry 
in \D4 (compare the discussion in footnote \ref{Ntwo}). In particular, 
the gauge fields are 
in \N2 vector multiplets, the untwisted matter fields in hyper multiplets 
and the twisted fields in hyper and vector-tensor multiplets. This constrains 
the possible couplings. The projection on \Z{N}-invariant states eliminates 
some of the fields (e.g. the the scalar partners of the gauge fields are 
projected out), but the remaining terms in the Lagrangian are inherited
from the \N2 theory. Now, it is important to note that in the \N2 \Z{m} 
theory all of the twisted fields in linear multiplets belong to vector-tensor
multiplets. In the \N1 \Z{N} theory some of the vectors are projected out,
but their linear partners still couple as required by \N2 supersymmetry.
In \cite{WLP} it was shown that in an \N2 supersymmetric theory only fields
belonging to vector multiplets can couple to the gauge kinetic terms.
Therefore a coupling $M^{(k)}\tr(W^{(a)}W^{(a)})$ as in (\ref{L_Mk})
is only possible if one either has only \N1 supersymmetry (i.e.\ no
fixed planes) or the $M^{(k)}$ belong to \N2 vectors, which means that
their duals $L^{(k)}$ are in \N2 vector-tensor multiplets.
We conclude, that besides the linear multiplets from the \N1 sectors all
the linear multiplets from the \N2 sectors, but none of the chiral
multiplets, can contribute to anomaly cancellation. Otherwise stated,
fields from all twisted sectors except the $k=N/2$ sector can contribute
to anomaly cancellation.\footnote{It is interesting to note that the 
coefficient $b_2^{(k,a)}$ in (\ref{bcoeff}) vanishes for $k=N/2$, 
because, for even $N$, $V_a$ is of the form $V_a={2j+1\over2N}$, 
with $j\in\bZ$ (see \cite{AFIV}).}
As we noted above these fields suffice to cancel 
all the $U(1)$ anomalies in \D4, \N1 type IIB orientifolds.

\section{K\"ahler anomalies}
In compact factorizable orbifolds of the heterotic string the tree-level
Lagrangian is invariant under $\prod_{i=1}^3SL(2,\RR)_i$ transformations 
acting on the K\"ahler moduli $T_i$ as
\be \label{def_SL}
T_i\ \to\ {a_iT_i-ib_i\over ic_iT_i+d_i},\qquad\w 
 \left(\begin{array}{cc}a_i &b_i\\ c_i &d_i\end{array}\right)\in SL(2,\RR)_i.
\ee
The K\"ahler potential for the $T_i$, $K^{(T)}=-\sum_i\ln(T_i+\bar T_i)$, 
transforms as
\be \label{K_transf}
K^{(T)}\ \to\ K^{(T)}+\ln(ic_iT_i+d_i)+\ln(-ic_i\bar T_i+d_i),
\ee
i.e.\ the kinetic terms of the $T_i$ are not modified ((\ref{K_transf}) 
is a K\"ahler transformation).
At one-loop one finds an anomalous variation of the Lagrangian (\ref{L_lnS})
\cite{LO,DFKZ,louis92}
\be \label{var_L}
\delta\cL^{S_{\rm dil}}_{\rm one-loop}={1\over32\pi^2}\ith
          \sum_{i=1}^3\bia\ln(ic_iT_i+d_i)\tr(W^{(a)}W^{(a)})\ +\ {\rm h.c.}
\ee
The anomaly coefficients $\bia$ are defined by
\be \label{def_bia}
\bia=-T(G_a)+\sum_{\cR_a}T(\cR_a)(1+2n^i_{\cR_a}),
\ee
where the sum is over all fields charged under $G_a$ (transforming in a 
representation $\cR_a$) and the modular weights $n^i_{\cR_a}$ can be 
read off from the K\"ahler potential of the charged fields $\Phi_r$, 
which to second order in $\Phi_r$ has the form
\be \label{Kmatter}
K^{\rm matter}=\sum_r\prod_{i=1}^3(T_i+\bar T_i)^{n^i_r}
               \Phi_r\bar \Phi_r.
\ee

However, the variation (\ref{var_L}) is cancelled by an opposite variation of 
$S_{\rm dil}$ like in the case of $U(1)$ gauge anomalies. The authors of 
\cite{AGNT} found that at one-loop the effective Lagrangian for the linear 
dilaton multiplet (\ref{L_lnL}) contains an additional term
\be \label{LlnT_term}
\ithth{1\over8\pi^2}\sum_{i=1}^3\tilde b^i_a\ln(T_i+\bar T_i)L_{\rm dil},
\ee
whith $\bia=k^{(a)}\tilde b^i_a$.
The dual Lagrangian (\ref{L_lnS}) is then modified to
\bea \label{L_lnSmod}
\cL^{S_{\rm dil}} &= &-\ithth b_1\ln\left(S_{\rm dil}+\bar S_{\rm dil}
                       -4b_3V^{(0)}-{1\over8\pi^2}\sum_{i=1}^3\tilde b^i_a
                                   \ln(T_i+\bar T_i)\right) \nonumber\\
   &&   +\qt\ith k^{(a)}S_{\rm dil}\tr(W^{(a)}W^{(a)})
        +\qt\ithb k^{(a)}\bar S_{\rm dil}\tr(\bar W^{(a)}\bar W^{(a)}).
\eea
We replaced $b_2^{(a)}$ by its value $-\half k^{(a)}$ (in units of 
$\alpha^\prime/2$). This leads to the transformation law
\be
S_{\rm dil}\ \to \ S_{\rm dil}-{1\over8\pi^2}\sum_{i=1}^3\tilde b^i_a
   \ln(ic_iT_i+d_i)
\ee
under (\ref{def_SL}) and cancels the anomaly. (For a detailed discussion
of K\"ahler anomalies and the restrictions they impose on heterotic string
vacua see \cite{IL92}).

The tree-level Lagrangian of \D4, \N1 type IIB orientifolds shows the same
symmetry under $\prod_{i=1}^3SL(2,\RR)_i$ transformations and again there 
are one-loop anomalies.
In \cite{IRU99} it has been proposed that the anomalies can be cancelled
by the same mechanism that cancels $U(1)$ gauge anomalies via exchange of
twisted fields $M^{(k)}$. To cancel also the K\"ahler anomalies these fields
need to transform under the K\"ahler transformations (\ref{def_SL}). In the
description in terms of linear multiplets this amounts to additional terms 
\be \label{LklnT_term}
\ithth{1\over8\pi^2}\sum_{i=1}^3\aik\ln(T_i+\bar T_i)L^{(k)}
\ee
in the Lagrangian (\ref{L_Lk}) such that
\be \label{bia_factor}
\sum_k2b_2^{(k,a)}\aik=-\bia.
\ee
The Lagrangian for the chiral multiplets (\ref{L_Mk}) is then modified to
\bea \label{L_Mkmod}
\cL^M &= &\ithth{1\over4b_1}\sum_k\left(M^{(k)}+\bar M^{(k)}
                          -4b_3^{(k)}V^{(0)} -{1\over8\pi^2}\sum_{i=1}^3
                            \aik\ln(T_i+\bar T_i)\right)^2 \\
   &&   -\hf\ith \sum_kb_2^{(a,k)}M^{(k)}\tr(W^{(a)}W^{(a)})
        -\hf\ithb \sum_kb_2^{(a,k)}\bar M^{(k)}\tr(\bar W^{(a)}\bar W^{(a)}),
   \nonumber
\eea
which leads to the transformation
\be \label{Mk_transf}
M^{(k)}\ \to \ M^{(k)}-{1\over8\pi^2}\sum_{i=1}^3\aik\ln(ic_iT_i+d_i)
\ee
under (\ref{def_SL}) and cancels the anomaly. Let us again make a remark
on mass dimensions. In our conventions $L^{(k)}$, $V^{(0)}$ and $T_i$
are dimensionless superfields, $M^{(k)}$, $b_1$, $b_3^{(k)}$ and $\aik$ 
have mass dimension 2, $b_2^{(k,a)}$ has dimension $-2$ and $W^{(a)}$ has 
dimension $3/2$.

It is interesting to introduce the composite $U(1)$ connection associated 
to K\"ahler trans\-for\-mations\footnote{A general definition of the composite
K\"ahler connection is given in \cite{WB}, chapter 23; the application to 
K\"ahler anomalies is discussed in \cite{LO,louis92}.}
\be \label{K_conn}
A^{(K)}_\mu=\qt\left({\partial K\over\partial T_i}\dm T_i
               -{\partial K\over\partial\bar T_i}\dm \bar T_i\right).
\ee
The field strength associated to this connection (with 
$K=-\sum_i\ln(T_i+\bar T_i)+\ldots$) is
\be \label{K_field}
F^{(K)}\mn=\hf{\dm\bar T_i\dn T_i-\dn\bar T_i\dm T_i\over(T_i+\bar T_i)^2}.
\ee
An expansion of (\ref{LklnT_term}) in components, eq.\ (\ref{lnPhiL_comp}), 
shows that it contains the term
\be
\ithth{1\over8\pi^2}\sum_{i=1}^3\aik\ln(T_i+\bar T_i)L^{(k)}
={1\over8\pi^2}\sum_{i=1}^3\aik(i\tilde F^{(K)}\mn B^{(k)\mu\nu})\ +\ \ldots,
\ee
where we used a partial integration for the term
$\tilde H^{(k)}_\mu\dM\I(T)/(T+\bar T)$. This result is just the analogue of 
the coupling (\ref{BF_term}) between the twisted RR-fields and the $U(1)_X$
gauge fields living on the D-branes, with $\aik i\tilde F^{(K)}\mn$ 
corresponding to $\tr(\gamma_k i\tilde F^{(0)}\mn)$ and $c_3=1/8\pi^2$. 
Up to factor of $-\alpha^\prime$, which is implicitly contained in $\aik$, 
this is the same normalization as the one chosen at the end of section 3.1.

In \cite{IRU99} it was shown that the anomaly coefficients $\bia$ indeed 
factorize as required in (\ref{bia_factor}):
\bea \label{bia_aik}
\bia &= &{1\over N}\sum_{k=0}^{N-1}\aikt\cos(4\pi kV_a) \\
\w &&\aikt=\left\{\begin{array}{ll}
            -C_k\tan(\pi kv_i) &\hbox{if $N$ odd}\\
            \eta_kC_{4k}\cot(2\pi kv_i)-C_k\cot(\pi kv_i)
               +\delta^i_3 2\eta_k {C_{4k}\over C_{2k}}\cos(2\pi kv_3)
                                       &\hbox{if $N$ even}
            \end{array}\right., \nonumber\\
          &&\eta_k=\left\{\begin{array}{ll} (-1)^k &{\rm if\ }N=6\\
              (-1)^{k/2} &{\rm if\ }N=12,\ k\ {\rm even}\\
              0 &{\rm if\ }N=12,\ k\ {\rm odd}
            \end{array}\right.,\qquad
            C_k=\prod_{i=1}^32\sin(\pi kv_i). \nonumber
\eea

For odd $N$ this can be transformed to
\be \label{bia_odd}
\bia={2\over N}\sum_{k=1}^{(N-1)/2}\tilde\alpha^i_{\bar k}\cos(2\pi kV_a),
\qquad\w \bar k=\left\{\begin{array}{ll} {k\over2} &{\rm if\ }k\ {\rm even}\\
              {N-k\over2} &{\rm if\ }k\ {\rm odd}
        \end{array}\right..
\ee
Comparison with (\ref{bia_factor}) gives
\be \label{aik_odd}
\aik={-\tilde\alpha^i_{\bar k}\over c_2C_kN}={\pm\tan(\pi\bar kv_i)\over c_2N},
\ee
where we used $b_2^{(a,k)}=c_2C_k\cos(2\pi kV_a)$ from (\ref{bcoeff})
and $C_k=\pm C_{\bar k}$ which is valid for all odd $N$ orientifolds.
The sign in (\ref{aik_odd}) is $+$ for \Z3 and \Z7, $k=2$, and $-$ 
for \Z7, $k=1,3$.

For even $N$ (\ref{bia_aik}) can be transformed to
\be \label{bia_even}
\bia={24\over N}\delta^i_3+
     {2\over N}\sum_{k=1}^{N/2-1}\aikt\cos(4\pi kV_a).
\ee
Two conclusions can be drawn from this expression: First, only for even $k$
is $\aik$ in (\ref{bia_factor}) \nonvan. Second, the anomalies corresponding
to the third complex plane cannot be cancelled by the exchange of twisted
fields $M^{(k)}$, because their coupling to the field strengths 
$\tr(W^{(a)}W^{(a)})$ is proportional to $\cos(2\pi kV_a)$, which is only
for $k=N$ independent of $V_a$ (one has $V_a={2j+1\over2N},j\in\bZ$). The
best way to compare the expression (\ref{bia_even}) for $\bia$ with
eq.\ (\ref{bia_factor}) is to consider each orientifold separately. We find
\bea
\bZ_6: && \bia= 4\delta^i_3-(2+2\delta^i_3)\cos(4\pi V_a), \nonumber\\
\bZ_6^\prime: && \bia=4\delta^i_3+(2+4\delta^i_2)\cos(4\pi V_a),  \nonumber\\
\bZ_{12}: && \bia=2\delta^i_3+{2\over3}\sqrt3(1-2\delta^i_2-\delta^i_3)
                    \cos(4\pi V_a)-(1+\delta^i_3)\cos(8\pi V_a). \nonumber
\eea
Comparing this to (\ref{bia_factor}), one obtains, for $i=1,2$:
\bea \label{aik_even}
\bZ_6: && \alpha^i_1=0,\ \alpha^i_2={-1\over3\sqrt3\,c_2}, \nonumber\\
\bZ_6^\prime: &&\alpha^i_1=0,\ \alpha^i_2={-(1+2\delta^i_2)\over\sqrt3\,c_2},\\
\bZ_{12}: && \alpha^i_1=0,\ \alpha^i_2={-(1-2\delta^i_2)\over3\,c_2},
             \ \alpha^i_3=0,\ \alpha^i_4={-1\over6\sqrt3\,c_2},\ \alpha^i_5=0.
         \nonumber
\eea
This shows that all K\"ahler anomalies, except those associated to the third
complex plane in even $N$ orientifolds, can indeed be cancelled by the 
mechanism proposed in \cite{IRU99}.
Whether the crucial coupling coupling (\ref{LklnT_term})
indeed occurs and whether the coefficients $\aik$ really have the
required values, should be verified by an explicit string calculation.

There is good reason to believe that the couplings (\ref{LklnT_term})
are generated with the correct coefficients. In four-dimensional heterotic
string vacua, there is an exact symmetry, called T-duality, which acts on 
the moduli $T_i$ as the K\"ahler transformation (\ref{def_SL}) but with 
integer coefficients $a_i,b_i,c_i,d_i$. The Green-Schwarz mechanism 
guarantees that this symmetry is preserved at the quantum level.
Because of heterotic-type I duality (see e.g.\ \cite{PW}) the same should 
be true for type IIB orientifolds. The type I moduli $T_i$ are defined by 
\cite{AFIV,IMR}
\be \label{def_TI}
T_i={R_i^2\over\lambda_I\alpha^\prime}+iC_{(2)2i+2,2i+3},
\ee
where $R_i$ is the radius of the $i$-th torus, $C_{(2)i,j}$ are the
internal components of the ten-dimensional RR 2-form and 
$\lambda_I=e^\varphi$ is the ten-dimensional type I string coupling.
Under the duality mapping given in \cite{PW} the type I $T_i$ are mapped
to the heterotic $T_i$, defined by $T_i=R_i^2/\alpha^\prime+iB_{2i+2,2i+3}$.
The heterotic T-duality should therefore be realized as a symmetry of
four-dimensional type I vacua (and in particular of type IIB orientifolds)
acting on the type I moduli $T_i$ as the $\prod_{i=1}^3SL(2,\bZ)_i$ subgroup
of (\ref{def_SL}). Note that on the type I side, this symmetry is not
T-duality.\footnote{T-duality in type I vacua exchanges different types
of D-branes. From the T-duality rules derived in \cite{IMR} one sees that
it also exchanges the three $T_i$ and the type I dilaton multiplet 
$S_{\rm dil}$. This is clearly different from (\ref{def_SL}).}

If D5-branes are present in the considered orientifold, the above reasoning
has to be slightly modified. The field strengths corresponding to the
gauge fields living on the D5-branes do not couple to $S_{\rm dil}$ but
rather to $T_3$ (here we assume that the D5-branes are extended in the four
non-compact dimensions and in the third complex plane). The corresponding
term in the Lagrangian is
\be \label{TWW_coupl}
\ith T_3\tr(W^{(a,5)}W^{(a,5)})\ +\ {\rm h.c.}
\ee
Clearly, the $SL(2,\bZ)_3$ subgroup of (\ref{def_SL}) is explicitly
broken by this coupling. Therefore one cannot expect the Green-Schwarz
mechanism to work for the K\"ahler symmetries associated to the third
complex plane if D5-branes are present. The existence of D5-branes in
type IIB orientifolds is intimately linked to the existence of an 
order-two twist. If the considered orientifold contains an order-two
twist, then tadpole cancellation requires D5-branes that are extended
in the complex plane which is fixed by this twist. Thus, whenever a 
complex plane is fixed under the action of $\theta^{N/2}$, then the
K\"ahler symmetry associated to this plane is explicitly broken.

There is a striking difference between the cancellation of K\"ahler
anomalies in heterotic and type I vacua. In the heterotic string,
only the anomalies in the \N1 sectors (i.e.\ the sectors without
fixed planes) are cancelled by a Green-Schwarz mechanism. The anomalies
in the \N2 sectors are cancelled by a $T_i$ dependent one-loop correction
to the gauge kinetic function. The authors of \cite{DKL} calculated the
one-loop correction $f^{(1)}$ to the gauge kinetic function $f$ (as defined
in (\ref{Lgen}), see \cite{LF} for a pedagogical presentation) for \D4, \N1 
heterotic orbifolds. They find
\be \label{f_one}
f^{(1)}={-1\over8\pi^2}\sum_{i=1}^3b^{\prime i(\cN=2)}_a\ln\eta^2(iT_i).
\ee
Because of the transformation property of the Dedekind $\eta$-function
under (\ref{def_SL}), $\eta^2(iT_i)\ \to\ (ic_iT_i+d_i)\eta^2(iT_i)$, 
the $\prod_{i=1}^3SL(2,\bZ)_i$ symmetry is restored at the quantum 
level.

In contrast to this, the gauge kinetic function of type I vacua can
only depend linearly on the $T_i$ \cite{ABD}. This can be understood
from the fact that the imaginary part of $T_i$ in eq.\ (\ref{def_TI}) 
is a RR-field, which leads to an invariance of the action under a
Peccei-Quinn symmetry $T_i\to T_i+ia_i$. As a consequence, the mechanism
for the cancellation of K\"ahler anomalies in the \N2 sectors in type
IIB orientifolds must be different from the mechanism that achieves 
anomaly cancellation in heterotic orbifolds. Indeed, we saw abow that
the generalized Green-Schwarz mechanism also works in the \N2 sectors
of type IIB orientifolds. The only orientifold that has fixed planes
under a twist $\theta^k$, with $k\neq N/2$, is the $\bZ_6^\prime$
orientifold. It would therefore be interesting to obtain the coupling
(\ref{LklnT_term}) for the $k=2$ sector of $\bZ_6^\prime$ from an 
explicit string calculation to see if the Green-Schwarz mechanism 
works in the expected way. If the K\"ahler anomalies associated to
the second complex plane of $\bZ_6^\prime$ are cancelled by the
mechanism described above, then the complete $SL(2,\RR)_2$ symmetry
is preserved at the quantum level. This differs from the situation
in the corresponding $\bZ_6^\prime$ orbifold of the heterotic string
where the classical $SL(2,\RR)_2$ symmetry is broken to $SL(2,\bZ)_2$
by quantum effects. Presumably the $SL(2,\RR)_2$ of the $\bZ_6^\prime$ 
orientifold is broken to its discrete subgroup by some \nonpert effect.

The authors of \cite{IRU99} find that there are also mixed 
K\"ahler-gravitational anomalies proportional to the coefficient
$b^i_{\rm grav}=b^i_{\rm cl}+b^i_{\rm op}$. The first contribution
is from closed string modes and the second from open string modes.
It turns out that the open string contribution can be cancelled 
by the twisted fields $M^{(k)}$ transforming as in (\ref{Mk_transf}).
But the $b^i_{\rm cl}$ contribution can only be cancelled by a 
transformation of the type I dilaton $S_{\rm dil}$ under
K\"ahler transformations:
\be \label{S_transf}
S_{\rm dil}\ \to \ S_{\rm dil}-{1\over8\pi^2}\sum_{i=1}^3b^i_{\rm cl}
                   \ln(ic_iT_i+d_i).
\ee
The K\"ahler potential of the dilaton is therefore modified to
\be \label{K_dil}
-\ithth \ln\left(S_{\rm dil}+\bar S_{\rm dil}
           -{1\over8\pi^2}\sum_{i=1}^3b^i_{\rm cl}\ln(T_i+\bar T_i)\right).
\ee
As noted in \cite{IRU99} the additional transformation of $S_{\rm dil}$ 
does not spoil the K\"ahler anomaly cancellation described above, because
the coefficient $b^i_{\rm cl}$ is of higher order than $\aik$ in the string 
loop expansion.

\section{Fayet-Iliopoulos terms}
The generalized Green-Schwarz mechanism described in the preceeding sections
cancels the gauge and K\"ahler anomalies, but it also gives mass to the
gauge bosons (of the anomalous $U(1)$'s), thus breaking gauge invariance. 
In addition, it generically leads to a non-trivial contribution to the 
D-term of the (pseudo-)anomalous vector field. This contribution is called 
Fayet-Iliopoulos term and it induces non-vanishing vacuum expectation values 
for some of the charged fields. (Fayet-Ilipoulos terms in type IIB orientifolds
have been discussed in \cite{CELW,LLN}.)

The Lagrangian describing the interactions of the dilaton $S_{\rm dil}$,
the untwisted moduli $T_i$, the twisted moduli $M^{(k)}$ and the gauge
fields of type IIB orientifolds is obtained from (\ref{L_Mkmod}), 
(\ref{TWW_coupl}) and (\ref{K_dil}):
\bea \label{total_L}
\cL &= &\ithth\Bigg(-\ln\left(S_{\rm dil}+\bar S_{\rm dil}
        -{1\over8\pi^2}\sum_{i=1}^3b^i_{\rm cl}\ln(T_i+\bar T_i)\right) \\
      &&\qquad+{1\over4b_1}\sum_k\left(M^{(k)}+\bar M^{(k)}
         -4b_3^{(k)}V^{(0)} -{1\over8\pi^2}\sum_{i=1}^3
         \aik\ln(T_i+\bar T_i)\right)^2\Bigg) \nonumber\\
      &&-\hf\ith \left(b^{(a)}_{2,\rm dil}S_{\rm dil}\ +\ 
         \sum_kb_2^{(a,k)}M^{(k)}\right)\tr(W^{(a,9)}W^{(a,9)})\ +\ {\rm h.c.}
      \nonumber\\
      &&-\hf\ith \left(b^{(a)}_{2,T}T_3\ +\ 
         \sum_kb_2^{(a,k)}M^{(k)}\right)\tr(W^{(a,5)}W^{(a,5)})\ +\ {\rm h.c.}
      \nonumber
\eea
The coefficient $b^{(a)}_{2,\rm dil}$ is defined by $\bar\cD^2L_{\rm dil}=
2b^{(a)}_{2,\rm dil}\tr(W^{(a,9)}W^{(a,9)})$. As in the low-energy effective
action of the heterotic string, this coefficient
is uniquely fixed by the supergravity algebra in \D{10}. In the conventions
of \cite{joe} one has $b^{(a)}_{2,\rm dil}=-\alpha^\prime/(2\sqrt2)$.
The indices 9 and 5 on the field strengths $W^{(a)}$ indicate on which type
of D-branes (9-branes or 5-branes) the corresponding gauge fields live.
The coupling of $T_3$ to the D5-branes cannot be understood from linear/chiral
multiplet duality, but its coefficient can be determined by perfoming a
T-duality transformation which interchanges D9-branes and D5-branes. This
yields $b^{(a)}_{2,T}=b^{(a)}_{2,\rm dil}$. Usually the chiral
dilaton multiplet is defined to be the coeffcient of $\qt\tr(WW)$. This is
obtained by redefining $S_{\rm dil}\to (\alpha^\prime/\sqrt2)S_{\rm dil}$.

From the results (\ref{FIterm}), (\ref{mass_term}) of appendix B, one easily 
finds the Fayet-Iliopoulos term $\xi^2_{\rm FI}$ and the gauge boson 
mass $m_A$ for an anomalous $U(1)_X$ embedded in $U(n_X)$ living on the
D9-branes (the result for a D5-brane $U(1)_X$ is obtained by replacing 
$S_{\rm dil}$ by $T_3$):
\bea  \label{xiFI}
\xi^2_{\rm FI} &= &-{g\over b_1}\sum_k b_3^{(k)}\left(
            M^{(k)}|+\bar M^{(k)}|
            -{1\over8\pi^2}\sum_{i=1}^3\aik\ln(T_i|+\bar T_i|)\right), \\
\label{mA}
m_A^2 &=&{4g^2\over b_1}\sum_k (b_3^{(k)})^2={4n_X^2c_3^2(N-\rho)g^2\over b_1},
\quad\w\rho=\left\{\begin{array}{ll}0&\hbox{if $N$ odd}\\2&\hbox{if $N$ even}
      \end{array}\right.,
\eea
where
\be \label{gaugefunction}
g^{-2}=\R(f)=\hf (S_{\rm dil}|+\bar S_{\rm dil}|)
               -\sum_kb_2^{(X,k)}(M^{(k)}|+\bar M^{(k)}|).
\ee
A vertical slash denotes the lowest component of a superfield.
In the second equality of (\ref{mA}) we replaced $b_3^{(k)}$ by
its value determined in (\ref{bcoeff}). The index $X$ refers to $U(1)_X$.
The coefficients $b_1$ and $c_3$ are both proportional to 
$\alpha^{\prime-1}=M_{\rm str}^2$. Thus, the (pseudo-)anomalous gauge
bosons have masses comparable in size to the typical masses found in 
heterotic orbifolds. In the normalization $b_1=\alpha^{\prime-1}$,
$c_3=-(8\pi^2\alpha^\prime)^{-1}$, we find
\be \label{mass_oddN}
m_A^2={f_1\over\pi^4}g^2M_{\rm str}^2
\ee
for the two odd $N$ orientifolds, with $f_1=27,\:7$ for \Z3, \Z7. The
Fayet-Iliopoulos term for \Z3 reads (in the same normalization)
\be \label{FI_Zthree}
\xi^2_{\rm FI}={3\sqrt3\over2\pi^2}g\left(M^{(1)}|+\bar M^{(1)}|
      -{\sqrt3\over4\pi^2\alpha^\prime}\sum_i\ln(T_i|+\bar T_i|)\right).
\ee

At the supersymmetric minimum the D-term (\ref{eom_D})
vanishes. As explained in the appendix B, this relates $\xi^2_{\rm FI}$
to the vacuum expectation values of the charged fields. If we assume
canonical kinetic terms and minimal coupling to the gauge fields,
$\bar\Phi_r e^{2q_rV}\Phi_r$, for the charged fields $\Phi_r$, 
then the Fayet-Iliopoulos term is given by
\be \label{eom_FI}
\xi^2_{\rm FI}=g\sum_rq_r|\phi_r|^2.
\ee
In most orientifold models there exist no fields which are charged under
$U(1)_X$ but neutral under the \nonab gauge group factors. Therefore
a \nonvan $\xi^2_{\rm FI}$ implies a breaking of the \nonab gauge symmetry
by the Higgs effect. As noted in \cite{LLN} this differs from the situation
in heterotic orbifolds, where \nonab singlets charged under $U(1)_X$
generically exist. A second difference to heterotic vacua is the possibility 
of having a vanishing Fayet-Iliopoulos in type IIB orientifolds \cite{IRU98}.
This is possible because of the dependence of $\xi^2_{\rm FI}$ on the
expectation value of the twisted fields. For type I vacua that are 
obtained by compactifying the ten-dimensional theory on a smooth
Calabi-Yau manifold instead of an orientifold there is no such dependence 
on twisted fields. Indeed, it has been shown in \cite{JMR} that the
Fayet-Iliopoulos in smooth type I vacua has exactly the same form as in
heterotic vacua. It is proportional to the universal anomaly coefficient
$\cA$ and given by $\xi^2_{\rm FI,CY}=\cA g^2M_{\rm str}^2/192\pi^2$.
Comparing this to the result for orientifolds, we see that the anomaly
coefficient $\cA$ is split into a sum of contributions from the different
twisted sectors, as in (\ref{cancel}).

If we insist on unbroken \nonab gauge symmetry, then in most orientifold
models we must require $\xi^2_{\rm FI}=0$. This relates the expectation 
value of the $M^{(k)}$ to $\sum_i\ln(T_i+\bar T_i)$. In the simplest
example of the \Z3 orientifold, one has from (\ref{FI_Zthree})
\be
M^{(1)}|+\bar M^{(1)}|=
      {\sqrt3\over4\pi^2\alpha^\prime}\sum_i\ln(T_i|+\bar T_i|).
\ee
The blowing-up of the orientifold, however, is related to the expectation
values of the lowest components of the linear multiplets $L^{(k)}$. From
the discussion in sections 3 and 4, it can be seen that in terms of
linear multiplets the Fayet-Iilopoulos term (\ref{xiFI}) is given
by $\xi^2_{\rm FI}=-{g\over b_1}\sum_k b_3^{(k)}L^{(k)}|$. Therefore,
a vanishing Fayet-Iliopoulos term corresponds to the orientifold limit.

\section{Conclusions and outlook}
We have seen that the duality between linear and chiral multiplets
provides an easy way to understand anomaly cancellation in \D4, \N1
type IIB orientifolds. Due to the existence of RR forms there appear
antisymmetric tensors (embedded in linear supermultiplets) in the 
twisted spectrum. Their coupling to the gauge fields on the D-branes
leads to a non-trivial transformation under the (pseudo-)anomalous
gauge symmetry of the scalars (embedded in chiral supermultiplets) which
are dual to these antisymmetric tensors. This transformation cancels the
anomaly. The crucial twist-number non-conserving couplings, which are
not present in heterotic orbifolds, are possible in type IIB orientifolds
because the corresponding world-sheets have boundaries and/or are 
non-orientable.

It is interesting to compare type IIB orientifolds to heterotic orbifolds.
There are several differences, a better understanding of which could
lead to a deeper insight into heterotic-type I duality \cite{CELW,LLN}. 
In \D4, \N1 heterotic orbifolds, $U(1)$ gauge anomalies are cancelled by a 
Green-Schwarz mechanism involving only one linear multiplet: the 
dilaton superfield. This is possible because of the universal nature 
of the anomalies. The non-universality of the anomalies in type IIB 
orientifolds is accounted for
by several linear multiplets that contribute to anomaly cancellation.
However their expectation values are not related to the string loop 
expansion but rather to the blowing-up of the orbifold singularities.
This has interesting phenomenological implications. Whereas in heterotic
orbifolds the value of the Fayet-Iliopoulos term is uniquely determined
once the expectation value of the dilaton is fixed, the same term in
type IIB orientifolds depends on the expectation value of the twisted
moduli and can even vanish in the orientifold limit. 

In heterotic orbifolds, the gauge kinetic function receives $T_i$-dependent
threshold corrections from the \N2 sectors. These are crucial to cancel
the anomalies in the discrete K\"ahler symmetries associated to fixed planes.
From general arguments such $T_i$-dependent threshold corrections cannot
exist in type IIB orientifolds, because of a Peccei-Quinn symmetry related
to the $T_i$ \cite{ABD}. As a consequence, the continuous $SL(2,\RR)$ 
K\"ahler symmetries associated to fixed planes of type IIB orientifolds 
are either completely broken by the anomaly or completely preserved by
the Green-Schwarz mechanism. In this article we showed that the Green-Schwarz
mechanism works even for the fixed planes (which are not fixed by an order-two
twist), but the precise coefficients of the required couplings have to be
determined by an explicit calculation. A break-down of the continuous
K\"ahler symmetries to their discrete subgroups seems to require some
\nonpert effect. In the \N2 sectors of heterotic orbifolds there are 
additional gauge group independent corrections to the gauge kinetic 
function \cite{WKLL,NilSti}. It would be interesting to see if such 
corrections are also present in type IIB orientifolds. By the argument
given above, they can however only depend on the $U_i$ but not on the $T_i$
to all orders in perturbation theory. 

The twisted spectrum of some type IIB orientifolds contains vector fields.
The corresponding gauge symmetry is inherited from the gauge invariance
of the RR forms. But it is not clear what is the role of this gauge symmetry
in \D4. In heterotic orbifolds this does not happen. There, all twisted
states are in chiral multiplets. Unfortunately, no heterotic dual is known
for the $\bZ_6^\prime$ or \Z{12} orientifolds, which possess twisted
vectors.

\vspace{1cm}

\begin{center} {\large\bf Acknowledgements} \end{center}
It is a pleasure to thank R.~Grimm for explaining many aspects of
linear/chiral multiplet duality to me, L.~Ib\'a\~nez for interesting
discussions and useful comments on the manuscript, and R.~Rabad\'an
for helping me to understand the technical details of the anomaly
cancellation mechanism. I would also like to thank M.~Haack, J.~Louis
and G.~Violero for valuable discussions.
This work is supported by a TMR network of the European Union, 
ref. FMRX-CT96-0090. 

\newpage
\begin{center}\huge\bf Appendix \end{center}
\begin{appendix}
In this appendix some useful results of \N1 supersymmetry are summarized. 
We use the notation and conventions of \cite{WB}.
\section{Linear Multiplets}
An \N1 linear multiplet $L$ is defined to be the most general real superfield
which satisfies the constraints
\be \label{lin_constr}
\cD^2L=\bar\cD^2L=0,
\ee
where
\bea \label{sderiv}
&&\cD_\alpha={\partial\over\partial\theta^\alpha}
       +i\sigma^\mu_{\alpha\dot\alpha}\bar\theta^{\dot\alpha}\dm, \qquad
  \bar\cD_{\dot\alpha}=-{\partial\over\partial\bar\theta^{\dot\alpha}}
            -i\theta^\alpha\sigma^\mu_{\alpha\dot\alpha}\dm,\\
&&\cD^2\equiv\cD^\alpha\cD_\alpha=-{\partial\over\partial\theta}
       {\partial\over\partial\theta}-2i\bar\theta\bar\sigma^\mu\dm
       {\partial\over\partial\theta}-\bar\theta\bar\theta\Box, \nonumber\\
&&\bar\cD^2\equiv\bar\cD_{\dot\alpha}\bar\cD^{\dot\alpha}=
       -{\partial\over\partial\bar\theta}{\partial\over\partial\bar\theta}
       -2i\theta\sigma^\mu\dm{\partial\over\partial\bar\theta}
       -\theta\theta\Box \nonumber
\eea
are the supercovariant derivatives. The solution of (\ref{lin_constr}) has
the component expansion
\bea \label{Lcomp}
L &= &l+\theta\chi+\bar\theta\bar\chi+\theta\sigma^\mu\bar\theta\tilde H_\mu
       +{i\over2}\theta\theta\dm\chi\sigma^\mu\bar\theta
       -{i\over2}\bar\theta\bar\theta\theta\sigma^\mu\dm\bar\chi
       -\qt\theta\theta\bar\theta\bar\theta\Box l, \\
  &&\w\ \dM\tilde H_\mu=0, \nonumber
\eea
where $l$ is a real scalar, $\chi$ a Weyl spinor, $\bar\chi$ its complex 
conjugate and $\tilde H_\mu$ a conserved vector. The constraint on 
$\tilde H_\mu$ can be satisfied by taking it to be the Hodge-dual of an 
antisymmetric tensor, $\tilde H_\mu=\hf\eps\dN B^{\rho\sigma}$.
For an arbitrary function $K(L)$ one finds
\bea \label{K_of_L}
\ithth K(L) &= &-\qt{\partial K\over\partial l}\myvrule{\theta=\bar\theta=0}
        \Box l\ -\ \qt
        {\partial^2K\over\partial l^2}\myvrule{\theta=\bar\theta=0}
        (\tilde H_\mu\tilde H^\mu -i\chi\sigma^\mu\dm\bar\chi 
         -i\bar\chi\bar\sigma^\mu\dm\chi) \nonumber\\
&&+\qt{\partial^3K\over\partial l^3}\myvrule{\theta=\bar\theta=0}
   \chi\sigma^\mu\bar\chi\tilde H_\mu\ +\ {1\over16}
   {\partial^4K\over\partial l^4}\myvrule{\theta=\bar\theta=0}
   \chi\chi\bar\chi\bar\chi. 
\eea
Thus, a quadratic K\"ahler potential for $L$ yields the usual kinetic terms
for $l$, $B\mn$ and $\chi$:
\be
-\ithth L^2=-\half\dm l\dM l\ +\ \half\tilde H_\mu\tilde H^\mu\ 
            -\ i\chi\sigma^\mu\dm\bar\chi\ 
            +\ {\rm total\ derivatives}.
\ee

To describe an antisymmetric tensor whose field strength is modified
by a Chern-Simons 3-form as in (\ref{Hdef}), 
$\dM\tilde H_\mu=\hf b_2\tr(F\MN\tilde F\mn)$, one needs a modified
linear multiplet, which satisfies
\be \label{modlin_constr}
\cD^2L=2b_2\tr(\bar W_{\dot\alpha}\bar W^{\dot\alpha}),\qquad 
\bar\cD^2L=2b_2\tr(W^\alpha W_\alpha),
\ee
where $W_\alpha$ is the chiral field strength multiplet corresponding
to the gauge field that is contained in the Chern-Simons form,
\bea \label{defW}
W_\alpha &= &-i\lambda_\alpha+\left(\delta_\alpha^{\ \beta}D-{i\over2}
          (\sigma^\mu\bar\sigma^\nu)_\alpha^{\ \beta}F\mn\right)\theta_\beta
          +\theta\sigma^\mu\bar\theta\dm\lambda_\alpha
          +\theta\theta\sigma^\mu_{\alpha\dot\alpha}
            \left(\dm\bar\lambda^{\dot\alpha}
          +i[A_\mu,\bar\lambda^{\dot\alpha}]\right) \nonumber\\
        &&-\hf\theta\theta(\sigma^\mu\bar\theta)_\alpha(i\dm D-\dN F\mn)
          -{i\over4}\theta\theta\bar\theta\bar\theta\Box\lambda_\alpha.
\eea
The constraint (\ref{modlin_constr}) can be solved by setting 
$L=L_0+b_2\Omega$, where $L_0$ is an unmodified linear multiplet
(i.e.\ $\cD^2L_0=\bar\cD^2L_0=0$) with component expansion as in 
(\ref{Lcomp}) and
\bea \label{defOmega}
\Omega &= &-2i\theta\sigma^\mu\tr(\bar\lambda A_\mu)
          -2i\bar\theta\bar\sigma^\mu\tr(\lambda A_\mu)
          +\hf\theta\theta\tr(\bar\lambda\bar\lambda)
          +\hf\bar\theta\bar\theta\tr(\lambda\lambda)
          +\theta\sigma^\mu\bar\theta\tilde Q_\mu \nonumber\\
       &&+\theta\theta\bar\theta\left(-i\tr(D\bar\lambda)
           -\bar\sigma^\mu\sigma^\nu 
           \tr\left(A_\nu(\dm\bar\lambda
                   +\mnorm{i\over2}[A_\mu,\bar\lambda])\right)
           +\tr(\bar\lambda\dM A_\mu)\right) \\
       &&+\bar\theta\bar\theta\theta\left(i\tr(D\lambda)
           -\sigma^\mu\bar\sigma^\nu
           \tr\left(A_\nu(\dm\lambda+\mnorm{i\over2}[A_\mu,\lambda])\right)
           +\tr(\lambda\dM A_\mu)\right) \nonumber\\
       &&+\theta\theta\bar\theta\bar\theta\left(-\hf\tr(D^2)+i\tr\left(\lambda
          \sigma^\mu(\dm\bar\lambda+i[A_\mu,\bar\lambda])\right)
          +\qt\tr(F\MN F\mn)\right). \nonumber
\eea
Here $\tilde Q_\mu$ is the Hodge dual of the Chern-Simons 3-form 
$Q_{\mu\nu\rho}=
 \tr(A_{[\mu}\dn A_{\rho]}-{2i\over3}A_{[\mu}A_{\nu}A_{\rho]})$,
and the traces are over the adjoint representation of the gauge group.
It is easy to check that
\bea
\bar\cD^2\Omega &= &-2\tr(\lambda\lambda)-4\theta\left(i\tr(D\lambda)+
       \hf\sigma^\mu\bar\sigma^\nu \tr(F\mn\lambda)\right)
       -4i\theta\sigma^\mu\bar\theta\tr(\lambda\dm\lambda) \nonumber\\
      &&+\theta\theta\left(2\tr(D^2)-4i\tr\left(\lambda\sigma^\mu
        (\dm\bar\lambda+i[A_\mu,\bar\lambda])\right)
        -\tr(F\MN F\mn)-i\tr(F\MN\tilde F\mn)\right) \nonumber\\
      &&+\theta\theta\partial_\rho\left(i\tr(\sigma^\mu\bar\sigma^\nu 
          F\mn\lambda)-2\tr(D\lambda)\right)\sigma^\rho\bar\theta 
          -\hf\theta\theta\bar\theta\bar\theta\Box\tr(\lambda\lambda)\\
      &= &2\tr(W^\alpha W_\alpha)   \nonumber
\eea
and analogously $\cD^2\Omega=2\tr(\bar W_{\dot\alpha} \bar W^{\dot\alpha})$. 
Consequently, the constraint (\ref{modlin_constr}) is satisfied.
The generalization of (\ref{K_of_L}) is given by
\bea \label{K_of_Lmod}
\hspace*{3mm}\ithth K(L) &= &\\
&&\hspace*{-2cm}-\qt{\partial K\over\partial l}\myvrule{\theta=\bar\theta=0}
        \bigg(\Box l+b_2\Big(2\tr(D^2)-4i\tr\left(\lambda\sigma^\mu
              (\dm\bar\lambda+i[A_\mu,\bar\lambda])\right)
        -\tr(F\MN F\mn)\Big)\bigg)\nonumber\\
&&\hspace*{-2cm}-\qt{\partial^2K\over\partial l^2}\myvrule{\theta=\bar\theta=0}
         \bigg(\tilde H_\mu\tilde H^\mu -i\hat\chi\sigma^\mu\dm\hat{\bar\chi} 
         -i\hat{\bar\chi}\sigma^\mu\dm\hat\chi
         -b_2^2\tr(\lambda\lambda)\tr(\bar\lambda\bar\lambda) \nonumber\\
&&\hspace*{-5mm}+b_2\hat\chi\Big(2i\tr(D\lambda)
                +\sigma^\mu\bar\sigma^\nu\tr(F\mn\lambda)\Big)\
                +b_2\hat{\bar\chi}\Big(2i\tr(D\bar\lambda)
         +\sigma^\mu\bar\sigma^\nu\tr(F\mn\bar\lambda)\Big)\bigg) \nonumber \\
&&\hspace*{-2cm}+\qt{\partial^3K\over\partial l^3}\myvrule{\theta=\bar\theta=0}
   \left(\hat\chi\sigma^\mu\hat{\bar\chi}\tilde H_\mu
   -{b_2\over2}\hat\chi\hat\chi\tr(\lambda\lambda) 
   -{b_2\over2}\hat{\bar\chi}\hat{\bar\chi}\tr(\bar\lambda\bar\lambda)\right)
   \quad+{1\over16}
   {\partial^4K\over\partial l^4}\myvrule{\theta=\bar\theta=0}
   \hat\chi\hat\chi\hat{\bar\chi}\hat{\bar\chi},\nonumber\\
&&{\rm where}\quad \hat\chi=\chi-2b_2i\sigma^\mu\tr(\bar\lambda A_\mu). 
  \nonumber
\eea

We also need the coupling of $L$ to an Abelian vector field $V^{(0)}$
and to chiral and antichiral fields $\Phi_i$, $\bar\Phi_i$. These have
the component expansion (in Wess-Zumino gauge)
\bea \label{Phicomp}
\Phi_i &= &\phi_i+\sqrt2\theta\psi_i+i\theta\sigma^\mu\bar\theta\dm\phi_i
         +\theta\theta F_i-{i\over\sqrt 2}\theta\theta\dm\psi_i\sigma^\mu
         \bar\theta+\qt\theta\theta\bar\theta\bar\theta\Box\phi_i,\\
\label{Vcomp}
V^{(0)} &= &-\theta\sigma^\mu\bar\theta A^{(0)}_\mu 
            +i\theta\theta\bar\theta\bar\lambda^{(0)} 
            -i\bar\theta\bar\theta\theta\lambda^{(0)} 
            +\half\theta\theta\bar\theta\bar\theta D^{(0)}.
\eea
We find
\bea \label{VL_comp}
\ithth V^{(0)}L &= &\hf A^{(0)\mu}\tilde H_\mu\ -\qt\hat\chi\sigma^\mu
               \hat{\bar\chi}A^{(0)}_\mu\ +{i\over2}\hat\chi\lambda^{(0)}
               \ -{i\over2}\hat{\bar\chi}\bar\lambda^{(0)}
               \ +\hf D^{(0)}l \\
          &= &\qt\tilde F^{(0)}\mn B\MN \ +\hf b_2A^{(0)\mu}\tilde Q_\mu
              -\qt\hat\chi\sigma^\mu \hat{\bar\chi}A^{(0)}_\mu\ 
              +{i\over2}\hat\chi\lambda^{(0)}\ 
              -{i\over2}\hat{\bar\chi}\bar\lambda^{(0)}\ 
              +\hf D^{(0)}l \nonumber
\eea
and
\bea \label{lnPhiL_comp}
\ithth \ln(\Phi+\bar\Phi)L &= &\ln(\phi+\bar\phi)
          \left(-\qt\Box l\ -\hf b_2\tr(D^2)\ +\qt b_2\tr(F\mn F\MN)\right) 
          \nonumber\\
        &&+{\dM\I(\phi)\tilde H_\mu\over\phi+\bar\phi}\ 
          -{F\bar F\over(\phi+\bar\phi)^2}l\ 
          +{\dm\phi\dM\bar\phi\over(\phi+\bar\phi)^2}l \\
        &&+\ {\rm fermionic}. \nonumber 
\eea

\section{D-terms of chiral models}
Consider a general Lagrangian depending on chiral and antichiral fields 
$\Phi_i$, $\bar\Phi_i$ and on an Abelian vector field V:
\bea \label{Lgen}
\cL &= &\ithth K(\Phi_i,\bar\Phi_i,V)+\qt\ith f(\Phi_i)W^\alpha W_\alpha
    +\qt\ithb f(\bar\Phi_i)\bar W_{\dot\alpha} W^{\dot\alpha} \nonumber\\
 && +\ith W(\Phi_i) +\ithb \bar W(\bar\Phi_i),
\eea
where the gauge kinetic function $f$ and the superpotential $W$ are
holomorphic functions of $\Phi_i$ and the K\"ahler potential $K$ is an
arbitrary real function. The component expansion of $\Phi_i$ and $V$ 
was given in (\ref{Phicomp}), (\ref{Vcomp}).
Expanding the Lagrangian (\ref{Lgen}) in components (see e.g.\ \cite{WB}),
\bea \label{Lgen_comp}
\cL &= &(-\dm\phi_i\dM\bar\phi_j\ +F_i\bar F_j)
 {\partial^2 K\over\partial\phi_i\partial\bar\phi_j}
 \myvruletwo{V=0}{\theta=\bar\theta=0} 
  \quad-\left(\I\left(A^\mu\dm\phi_i{\partial\over\partial\phi_i}\right)
  -\hf D\right){\partial\over\partial V}K
  \myvruletwo{V=0}{\theta=\bar\theta=0} \nonumber\\
&&-\qt A_\mu A^\mu {\partial^2\over\partial V^2}K
  \myvruletwo{V=0}{\theta=\bar\theta=0}\quad
  -\qt\R(f(\phi_i))F\mn F\MN\quad+\qt\I(f(\phi_i))F\mn\tilde F\MN
  \nonumber \\
&&+\hf\R(f(\phi_i))D^2\quad+{\partial\over\partial\phi_i}W
  \myvrule{\theta=\bar\theta=0}F_i\quad+{\partial\over\partial\bar\phi_i}\bar W
  \myvrule{\theta=\bar\theta=0}\bar F_i \\
&&+{\rm\ fermionic}, \nonumber
\eea
one finds the equations of motion for the auxiliary fields $F_i$, $D$:
\be \label{eom}
{\partial^2 K\over\partial\phi_i\partial\bar\phi_j}
\myvruletwo{V=0}{\theta=\bar\theta=0}\bar F_j=
-{\partial\over\partial\phi_i}W\myvrule{\theta=\bar\theta=0},
\qquad D={-1\over2\R(f(\phi_i))}{\partial\over\partial V}K
  \myvruletwo{V=0}{\theta=\bar\theta=0}.
\ee

We are interested in the case where one of the fields (say $\Phi_0$) appears
as an axion like in the Green-Schwarz mechanism and the other fields 
($\Phi_i,i\neq0$) couple to $V$ in the usual way,
\be
K=K(\Phi_0+\bar\Phi_0+4b_3V,\bar\Phi_ie^{2q_iV}\Phi_i), \qquad
f=f(\Phi_0).
\ee
Substituting into (\ref{Lgen_comp}) we find the component Lagrangian 
\cite{Quev}
\bea \label{GSLagr}
\cL &= &-K^{\prime\prime}\,\Big(\dm\R(\phi_0)\dM\R(\phi_0)
        +(\dm\I(\phi_0)+2b_3A_\mu)(\dM\I(\phi_0)+2b_3A^\mu)\Big) \\
&&-K_{i\jb}\,(\dm+iq_iA_\mu)\phi_i\,(\dM-iq_iA^\mu)\bar\phi_j+2b_3DK^\prime
   +\sum_iq_iDK_i\phi_i+\half\R(f(\phi_0))D^2 \nonumber\\[-1ex]
&&-\quarter\R(f(\phi_0))F\mn F\MN+\quarter\I(f(\phi_0))F\mn\tilde F\MN
  +K_{i\jb}F_i\bar F_j+W_iF_i+\bar W_i\bar F_i,\nonumber\\
&&+{\rm\ fermionic}, \nonumber
\eea
where a prime denotes the derivative with respect to $\phi_0$ evaluated at
$V=0=\theta=\bar\theta$ and an index $i$ on $K$ or $W$ denotes the derivative 
with respect to $\phi_i$. The equation of motion for $D$ now reads
\be \label{eom_D}
D={-1\over\R(f(\phi_0))}(2b_3K^\prime+\sum_iq_iK_i\phi_i).
\ee
Comparison of (\ref{GSLagr}) with a Lagrangian of the form (\ref{L_phi})
relates $f(\phi_0)$ to the gauge coupling constant, $g=f(\vev{\phi_0})^{-\hf}$.
To obtain the physical gauge fields, one has to rescale the vector field:
$V\to gV$.\footnote{More precisely, we write $V=g\hat V$ and express the
whole Lagrangian in terms of the physical field $\hat V$.}
The vacuum expectation value of the first term in the rescaled expression
for $D$, eq.\ (\ref{eom_D}), is called the Fayet-Iliopoulos term:
\be \label{FIterm}
\xi^2_{\rm FI}={-2b_3K^\prime\over\sqrt{\R(f(\vev{\phi_0}))}}.
\ee
(The square indicates that it has mass dimension 2.) At the supersymmetric
minimum $D$ vanishes, and therefore some of the charged fields acquire
vacuum expectation values,
$\xi^2_{\rm FI}=\R(f)^{-\hf}\sum_iq_iK_i\vev{\phi_i}$.

Another interesting fact is that (\ref{GSLagr}) contains a mass term for
the gauge fields. After rescaling one finds
\be \label{mass_term}
m_A^2={8b_3^2K^{\prime\prime}\over\R(f(\vev{\phi_0}))}.
\ee

\end{appendix}

\end{document}